\documentclass[acmsmall,screen,nonacm]{acmart}
\pdfoutput=1

\usepackage{cleveref}
\usepackage{xspace}
 
\usepackage[utf8]{inputenc}
\usepackage{subfiles}
\usepackage{comment}
\usepackage{mathtools}
\usepackage{inputenc}
\usepackage{microtype}
\usepackage{subcaption}
\usepackage{xcolor}
\usepackage{booktabs,longtable}
\usepackage{array}
\usepackage{pbox}
\usepackage{appendix}
\usepackage[mode=buildnew]{standalone}%
\usepackage{tikz}
\usepackage{graphicx}
\usepackage{amsthm}
\usepackage{mathrsfs}
\usepackage{paralist}
\usepackage{bussproofs}
\usepackage{siunitx}
\EnableBpAbbreviations

\usepackage{bm}

\crefname{equation}{}{}
\crefname{condition}{condition}{conditions}
\crefname{item}{}{}
\crefname{section}{Chapter}{Chapters}
\crefname{subsection}{Section}{Sections}

\definecolor{mypurple}{rgb}{0.5, 0.0, 0.5}

\definecolor{effectivenessCurve}{RGB}{151, 4, 12}
\definecolor{yel}{RGB}{237, 164, 60}
\definecolor{CostCurve}{RGB}{5, 20, 160}

\newcommand{\ca}{\mathbf}
\newcommand{\inp}{\mathrm{in}}
\newcommand{\out}{\mathrm{out}}

\newcommand{\RR}{\mathbb{R}}

\usetikzlibrary{positioning,decorations.pathreplacing,decorations.pathmorphing,shapes.arrows,decorations.markings,arrows.meta,calc,fit,quotes,cd,math,decorations.markings,arrows,backgrounds,shapes.geometric,shapes}

\newcommand{\ncat}[1]{\mathbf{#1}} %

\newcommand{\Set}{\ncat{Set}}

\newcommand{\Cat}{\ncat{Cat}}

\newcommand{\Lin}{\ncat{Lin}}

\newcommand{\SmallBox}[3]
{\begin{tikzpicture}[oriented WD, baseline=-2.5pt, bb Small]
\node[inner sep=.1cm] [bb={1}{1}] (X) {$\scriptstyle #3$};
\draw[label] node[left=.1 of X_in1] (Y) {$#1$}
             node[right=.1 of X_out1] {$#2$};
\end{tikzpicture}}

\newcommand{\SmallBoxTwo}[4]
{\begin{tikzpicture}[oriented WD, baseline=-2.5pt, bb Small]
\node[inner sep=.1cm] [bb={2}{1}] (X) {$\scriptstyle #4$};
\draw[label] node[left=.1 of X_in1] (Y) {$#1$}
node[left=.1 of X_in2] (Y) {$#2$}
node[right=.1 of X_out1] {$#3$};
\end{tikzpicture}}

\newcommand{\SmallBoxThree}[5]
{\begin{tikzpicture}[oriented WD, baseline=-2.5pt, bb Small]
\node[inner sep=.1cm] [bb={3}{1}] (X) {$\scriptstyle #5$};
\draw[label] node[left=.1 of X_in1] (Y) {$#1$}
node[left=.1 of X_in2] (Y) {$#2$}
node[left=.1 of X_in3] (Y) {$#3$}
node[right=.1 of X_out1] {$#4$};
\end{tikzpicture}}

\tikzcdset{%
   arrow style=tikz,
   diagrams={>={Classical TikZ Rightarrow[angle=63:4pt, line width=.6pt]}},
   arrows={semithick}
}

\tikzset{
  tick/.style={postaction={
    decorate,
    decoration={markings, mark=at position 0.5 with {\draw[-] (0,.4ex) -- (0,-.4ex);}}}
  },
  tickx/.style={
    postaction={ decorate,
      decoration={markings,
        mark=at position 0.5 with {
          \fill circle [radius=.28ex];
        }
      }
    }
  }
}
\newcommand{\tickar}
{\begin{tikzcd}[baseline=-0.5ex,cramped,sep=small,ampersand replacement=\&]{}\ar[r,tick]\&{}\end{tikzcd}}
\tikzset{
   dom/.style={append after command={coordinate[alias=dom#1]}},
   domA/.style={dom=A}, domB/.style={dom=B},
   domC/.style={dom=C}, domD/.style={dom=D},
   domE/.style={dom=E}, domF/.style={dom=F},
   cod/.style={append after command={coordinate[alias=cod#1]}},
   codA/.style={cod=A}, codB/.style={cod=B},
   codC/.style={cod=C}, codD/.style={cod=D},
   codE/.style={cod=E}, codF/.style={cod=F}
}

\tikzset{
   oriented WD/.style={%
      every to/.style={out=0,in=180,draw},
      label/.style={
         font=\everymath\expandafter{\the\everymath\scriptstyle},
         inner sep=0pt,
         node distance=2pt and -2pt},
      semithick,
      node distance=1 and 1,
      decoration={markings, mark=at position .5 with {\arrow{stealth};}},
      ar/.style={postaction={decorate}},
      execute at begin picture={\tikzset{
         x=\bbx, y=\bby,
         every fit/.style={inner xsep=\bbx, inner ysep=\bby}}}
      },
   bbx/.store in=\bbx,
   bbx = 1.5cm,
   bby/.store in=\bby,
   bby = 1.75ex,
   bb port sep/.store in=\bbportsep,
   bb port sep=2,
   bb port length/.store in=\bbportlen,
   bb port length=4pt,
   bb min width/.store in=\bbminwidth,
   bb min width=1cm,
   bb rounded corners/.store in=\bbcorners,
   bb rounded corners=2pt,
   bb small/.style={bb port sep=1, bb port length=2.5pt, bbx=.4cm, bb min width=.4cm, bby=.7ex},
   bb Small/.style={bb port sep=1, bb port length=2.5pt, bbx=.5cm, bb min width=.5cm, bby=1ex},
   bb/.code 2 args={%
      \pgfmathsetlengthmacro{\bbheight}{\bbportsep * (max(#1,#2)+1) * \bby}
      \pgfkeysalso{draw,minimum height=\bbheight,minimum width=\bbminwidth,outer sep=0pt,
         rounded corners=\bbcorners,thick,
         prefix after command={\pgfextra{\let\fixname\tikzlastnode}},
         append after command={\pgfextra{\draw
            \ifnum #1=0{} \else foreach \i in {1,...,#1} {
               ($(\fixname.north west)!{\i/(#1+1)}!(\fixname.south west)$) +(-\bbportlen,0) coordinate
               (\fixname_in\i) -- +(\bbportlen,0) coordinate (\fixname_in\i')}\fi %
            \ifnum #2=0{} \else foreach \i in {1,...,#2} {
               ($(\fixname.north east)!{\i/(#2+1)}!(\fixname.south east)$) +(-\bbportlen,0) coordinate
               (\fixname_out\i') -- +(\bbportlen,0) coordinate (\fixname_out\i)}\fi;
         }}}
   },
   bb name/.style={append after command={\pgfextra{\node[anchor=north] at (\fixname.north) {#1};}}}
}

 \definecolor{composite}{RGB}{151, 4, 12}

\newcommand{\Hair}{\ifmmode\mskip1mu\else\kern0.08em\fi}

\acmBooktitle{}
\begin{document}

\title{Categorical Semantics of Cyber-Physical Systems Theory}

\author{Georgios Bakirtzis}
\orcid{0000-0003-4992-0193}
\email{bakirtzis@virginia.edu}
\affiliation{%
  \institution{University of Virginia}
  \country{USA}
}

\author{Cody H. Fleming}
\orcid{0000-0001-6335-471X}
\email{flemingc@iastate.edu}
\affiliation{%
  \institution{Iowa State University}
  \country{USA}
}

\author{Christina Vasilakopoulou}
\email{cvasilak@math.upatras.gr}
\affiliation{%
  \institution{University of Patras}
  \country{Greece}
}

\begin{abstract}
  Cyber-physical systems require the construction and management of various models to assure their correct, safe, and secure operation. These various models are necessary because of the coupled physical and computational dynamics present in cyber-physical systems.
However, to date the different model views of cyber-physical systems are largely related informally, which raises issues with the degree of formal consistency between those various models of requirements, system behavior, and system architecture.
  We present a category-theoretic framework to make different types
  of composition explicit in the modeling and analysis of cyber-physical systems, which could assist in verifying the system as a whole.
  This compositional framework for cyber-physical systems gives rise to unified system models, where system behavior is hierarchically decomposed and related to a system architecture using the systems-as-algebras paradigm.
  As part of this paradigm, we show that an algebra of (safety) contracts generalizes over the state of the art, providing more uniform mathematical tools for constraining the behavior over a richer set of composite cyber-physical system models, which has the potential of minimizing or eliminating hazardous behavior.%
\end{abstract}


\maketitle

\pagestyle{plain} 
\section{Introduction}

In this paper we study the problem of unification
between the disparate but necessary models
used to assure correctness in cyber-physical systems (\textsc{cps}),
including requirements, system behaviors, and system architectures.
Currently, these views are largely managed
in an informal, piecemeal fashion, with no notion
of formal traceability between different \emph{types}
of models, which could lead to designing and implementing systems that are ultimately unsafe due to inconsistencies between these three views.
Model-based design attempts
to address some of the aforementioned issues,
but while it contains a notion of formal composition within each model view, it lacks a notion of formal composition
\emph{between} different model types.

This need for formal compositional theories
to support the design of \textsc{cps} is a consistent theme in \textsc{cps} literature both in the general modeling sense~\cite{bliudze:19} and particularly in contract-based design~\cite{benveniste:2018,nuzzo:2018}.
First, the model-based design of \textsc{cps} can be greatly assisted by the composition of different \emph{types} of models, which would provide traceability between the coupled physical and computational dynamics present in \textsc{cps}~\cite{allgower:2019}.
Second, the application of formal composition makes precise abstraction and refinement, which are necessary in model-based design and analysis~\cite{rungger:2014}.
Third, by investing in a compositional modeling paradigm we are better able to identify unsafe or uncontrolled interactions between subsystems~\cite{rushby:2011}.
We posit that category theory
and, specifically, the wiring diagram formalism~\cite{spivak:2016} provide an appealing framework
to build and analyze compositional models of \textsc{cps}.

In the design and analysis of \textsc{cps}, the word \emph{composition} appears in many different contexts and may refer to different things. Category theory is one context where its meaning is formal and refers to something specific, namely the partial operation on morphisms of a category. However, we will herein also occasionally use the term \emph{composite system} in its more relaxed sense, which we formalize categorically in this work  using the systems-as-algebras framework. On the contrary, the term \emph{compositionality} is not a formal one, rather the general characteristic of an analysis that ensures that the behavior of the whole is determined by the behavior of its building blocks.%

Wiring diagrams are a particularly interesting example
of the congruence between category theory
and model-based design.
Wiring diagrams have been independently created
by category theorists \cite{OperadofWiringDiagrams,Rupel.Spivak:2013a,Vagner.Spivak.Lerman:2015a} but surprisingly look and \emph{feel}
similar to engineering block diagrams used as the basic diagrammatic framework
for modeling, for example, the unified modeling language (\textsc{uml}), the systems modeling language (SysML), the generic modeling environment (\textsc{gme}), and a variety of tools from Mathworks including Simulink.
These types of diagrams are increasingly part of various research directions in \textsc{cps}, for example the Ptolemy project~\cite{buck:2001} or Möbius~\cite{masetti:2018}.
Systems engineering is a discipline where diagrammatic reasoning
has long been considered an important element
in managing complexity.
But several challenges persist, for example using SysML
for the analysis of systems designs means a scarcity
of simulation capabilities, an increased modeling effort to capture different views of the system,
and the need to maintain all these differing views concurrently even as they evolve asynchronously.
While the approach using wiring diagrams has little tool support currently, as an intellectual framework they overcome these limitations
by augmenting this diagrammatic reasoning
with stronger mathematical semantics.

In general, categorical semantics avoid
modeling the internal structure of the objects they act upon. Instead, an object is perceived through its relationships with other objects and not -- as is common with systems models -- by what the object is individually.
Indeed, in this context we focus on abstraction,
which we see as determining \emph{only} what is essential
in each layer of a given model.
This allows us to talk about how things are \emph{related}
instead of focusing on how things \emph{are}.
This mindset as applied to systems theory gives rise
to a circumspection of the system
where we do not examine a system by its individual elements
but by looking at the compositional structure of the system as a whole.
This might sound familiar to safety experts where it is -- arguably -- accepted 
that we cannot examine how safe a system is by examining its individual constituents~\cite{leveson:2011}.
Instead, by modeling \textsc{cps} in the wiring diagram framework we examine the system both
by the individual constituents and their specific interconnections, compositionally.

\emph{Contributions}. In this paper we use categories as a unifying modeling language for \textsc{cps}:
\begin{itemize}
\item We develop a categorical semantics of compositional \textsc{cps} theory that has the capability to merge physical models with computational models for the design and analysis of \textsc{cps}.\Hair\footnote{Compositional \textsc{cps} theory is a flavor of what Lee calls computational dynamical systems theory~\cite{lee:2006}.}
\item We formalize the general diagrammatic syntax of boxes and wires by adapting the systems-as-algebras model~\cite{spivak:2016} for \textsc{cps}, thereby producing a formal diagrammatic language for the design and analysis of \textsc{cps}.
\item We establish that the categorical and diagrammatic syntax equipped with a contracts algebra generalizes over the current state of the art~\cite{Contractsforsystemdesign}.
\end{itemize}

As diagrammatic reasoning takes an increasingly central role
in the modeling, simulation, and development
of \textsc{cps}, such relational semantics will become important
in type checking, navigating different domains of abstraction,
and ultimately assisting with providing evidence
that \textsc{cps} operate correctly during deployment.
This requires an effort both from industry and academia
to accept that visualization (usually the domain of industry)
\emph{and} mathematical rigor (usually the domain of academia) will be necessary
to improve the current state of the art
in system design.
Wiring diagrams are one answer to this merger
by implementing formal diagrammatic reasoning
for \textsc{cps} modeling and analysis.

\section{Categorical Background}

In this section we present some essential categorical machinery that will be used
to build up a formal compositional \textsc{cps} theory.

\subsection{A Few Basic Categorical Concepts}\label{sec:basiccats}

Briefly, a \emph{category} $\ca{C}$ consists 
of a collection of objects $X, Y, \ldots, Z$ and a collection 
of arrows $f\colon X\to Y$, along with a composition rule 
\begin{displaymath}
(f\colon X\to Y,g\colon Y\to Z)\mapsto g\circ f\colon X\to Z
\end{displaymath}
and an identity arrow $1_X\colon X\to X$ for all objects, subject to associativity and unity conditions: $(f\circ g)\circ h=f\circ (g\circ h)$ and $f \circ 1_X=f=1_Y\circ f$.
This definition encompasses a vast variety of structures 
in mathematics and other sciences: to name a few, $\mathbf{Set}$ is the category 
of sets and functions, whereas $\mathbf{Lin}$ is the category 
of $k$-linear (vector) spaces and $k$-linear maps between them, and we also have the category
of states and transitions between them~\cite{diskin:2015}.
For a complete treatment of basic categorical concepts, consult Lawvere
and Schanuel~\cite{lawvere:2009}, Leinster~\cite{leinster}, or Spivak~\cite{spivak:2014}. %

A standard diagrammatic way to express composites is $X\xrightarrow{f}Y\xrightarrow{g}Z$
and equations via commutative diagrams of the following form
\begin{displaymath}
\begin{tikzcd}
X\ar[r,"1"]\ar[dr,"f"'] & X\ar[d,"f"]\\
& Y
\end{tikzcd}\quad\textrm{ stands for }f\circ1_X=f
\end{displaymath}
A morphism $f\colon X\to Y$ is called \emph{invertible} or an \emph{isomorphism} when there exists another $g\colon Y\to X$ such that $f\circ g= 1_Y$ and $g\circ f=1_X$.

A \emph{functor} $F\colon\ca{C}\to\ca{D}$ between two categories consists of a function between objects and a function between morphisms, where we denote $Ff\colon FX\to FY$, such that it preserves composition and identities: $F(f\circ g)=Ff\circ Fg$ and $F(1_X)=1_{FX}$.
A functor can informally be thought of as a structure-preserving map between domains of discourse.
Interestingly, categories and functors form a category on their own,
denoted $\Cat$, in the sense that functors compose and the rest of the axioms hold.

A \emph{monoidal} category $\ca{V}$ is a category that comes equipped with a functor called `tensor product'
\begin{displaymath}
\otimes\colon\ca{V}\times\ca{V}\to\ca{V}
\end{displaymath}
which can be thought of as multiplication of objects and morphisms, or more broadly as doing operations in parallel. The tensor product comes with invertible morphisms $(X\otimes Y)\otimes Z\cong X\otimes (Y\otimes Z)$ meaning that it is associative up to isomorphism. %
There is also a distinguished object $I\in\ca{V}$ with $I\otimes X\cong X\cong X\otimes I$, acting like an identity for this multiplication. All these data satisfy certain axioms found, for example, in Joyal and Street~\cite{BraidedTensorCats}, that are beyond the scope of this paper.

Widely used examples of monoidal categories include $(\mathbf{Set},\times,\{*\})$ 
with the cartesian product of sets and the singleton,
as well as $(\mathbf{Lin},\otimes_k,k)$ with the tensor product of $k$-vector spaces.
Moreover, $(\ca{Cat},\times,\ca{1})$ with the cartesian product of categories (similarly to that of sets)
and the unit category with a single object and single arrow forms a monoidal category.
In fact, all these are examples of \emph{symmetric} monoidal categories, 
which come further equipped with isomorphisms $X\otimes Y\cong Y\otimes X$, 
for example, for two sets $X\times Y\cong Y\times X$ via the mapping $(x,y)\mapsto (y,x)$.

A \emph{lax monoidal} functor between two monoidal categories $F\colon(\ca{V},\otimes_\ca{V},I_\ca{V})\to(\ca{W},\otimes_\ca{W},I_\ca{W})$ is a functor that preserves the monoidal structure in a lax sense (meaning not up to isomorphism). Explicitly, it comes equipped with collections of morphisms, the `laxator' $\phi_{X,Y}\colon FX\otimes_\ca{W} FY\to F(X\otimes_\ca{V} Y)$ and the `unitor' $\phi_0\colon F(I_\ca{V})\to I_\ca{W}$ that express the relation between the image of the tensor and the tensor of the images inside the target category $\ca{W}$; these also adhere to certain axioms \cite{BraidedTensorCats}. Monoidal categories and lax monoidal functors also form a category of their own, denoted $\textbf{MonCat}_\textrm{lax}$.

\subsection{The Category \texorpdfstring{$\ca{W}$}{\textbf{W}} of Wiring Diagrams}\label{sec:categoryW}

The cornerstone of this work is the category $\ca{W}$ of \emph{labeled boxes} 
and \emph{wiring diagrams}. %
Informally, the objects of this category are to be thought of as 
empty placeholders for processes, so far only specifying the types of the input and output data that they may receive. For example, an object $X$ with inputs being pairs of a real and a natural number and outputs true or false values is diagrammatically depicted as
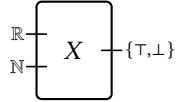
\begin{displaymath}
\begin{tikzpicture}[oriented WD, bbx=.1cm, bby =.1cm, bb port sep=.15cm]
\node [bb={2}{1}] (X) {$X$};
\draw[label]
node[left=.1 of X_in1]  {$\mathbb{R}$}
node[left=.1 of X_in2]  {$\mathbb{N}$}
node[right=.1 of X_out1]  {$\{\top,\bot\}$};
\end{tikzpicture}
\end{displaymath}
The two input wires above can also be represented by a single wire typed $\mathbb{R}\times\mathbb{N}$.
A process that can later be positioned inside that box is, for example, the function $$f(r,n)=\begin{cases}
\top & \textrm{if } r=n \\
\bot & \textrm{if }r\neq n
\end{cases}$$ To begin with, however, these boxes are uninhabited: they merely represent the architecture of a possible system. 

These interfaces, with finitely many input and output wires along with their associated types, are essentially the building blocks for forming larger interfaces from smaller ones, and this is what is captured by the morphisms in the category $\ca{W}$. For example, suppose $\SmallBox{\mathbb{C}}{\mathbb{R}}{Y}$ is another box. Intuitively, since the type of the output wire of $Y$ matches the type of one of the input wires of $X$, they could be linked along that wire
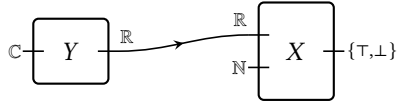
\begin{displaymath}
\begin{tikzpicture}[oriented WD, bbx=.1cm, bby =.1cm, bb port sep=.15cm]
\node [bb={1}{1}] (Y) {$Y$};
\node [bb={2}{1}, right=2cm of Y] (X) {$X$};
\draw[label]
node[above left= and .1 of X_in1]  {$\mathbb{R}$}
node[left=.1 of Y_in1]  {$\mathbb{C}$}
node[left=.1 of X_in2]  {$\mathbb{N}$}
node[right=.1 of X_out1]  {$\{\top,\bot\}$}
node[above right= and .1 of Y_out1]  {$\mathbb{R}$};
\draw[ar] (Y_out1) to (X_in1);
\end{tikzpicture}
\end{displaymath}
to provide a new interface that receives two inputs, one complex and one natural number, and outputs a true or false: 
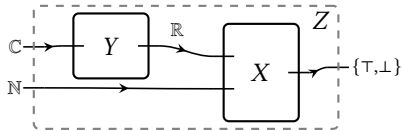
\begin{displaymath}
\begin{tikzpicture}[oriented WD, bbx=.1cm, bby =.1cm, bb port sep=.15cm]
\node [bb={1}{1}] (Y) {$Y$};
\node [bb={2}{1}] at (20,-3.5) (X) {$X$};
\node [bb={2}{1},fit=(X)(Y),draw,gray,dashed,inner xsep=15pt] (Z) {};
\draw[label]
node[above left=.1in and .15in of X_in1]  {$\mathbb{R}$}
node[left=.1 of Z_in1]  {$\mathbb{C}$}
node[left=.1 of Z_in2]  {$\mathbb{N}$}
node[right=.1 of Z_out1]  {$\{\top,\bot\}$};
\draw[ar] (Y_out1) to (X_in1);
\draw[ar] (Z_in1') to (Y_in1);
\draw[ar] (Z_in2') to (X_in2);
\draw[ar] (X_out1) to (Z_out1');
\node at (28,3.4) {$Z$};
\end{tikzpicture}
\end{displaymath}
While combining interfaces together, we want to be able to express not only the new interface they form, which in the above example is  $\SmallBoxTwo{\mathbb{C}}{\mathbb{N}}{\{\top,\bot\}}{Z}$, but also keep track of the intermediate wires. In our envisioned category, this will be expressed as a morphism from `$X$ and $Y$' into $Z$.\Hair\footnote{This morphism is explicitly given later in \cref{sec:contracts}.}

\begin{definition}
There is a category $\ca{W}$ with pairs of sets $X=(X_\inp,X_\out)$ as objects, thought 
of as the products of types of the input and output ports of an empty box as in
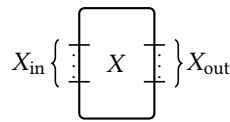
\begin{displaymath}
\begin{tikzpicture}[oriented WD, bbx=.1cm, bby =.1cm, bb port sep=.17cm]
\node [bb={2}{2}] (X) {$X$};
\draw[label]
node at ($(X.east)+(.7,.7)$) {$\vdots$}
node at ($(X.west)+(-.7,.7)$) {$\vdots$}
node[above right= of X_out1] (a) {}
node[below right= of X_out2] (b) {}
node[above left= of X_in1] (c) {}
node[below left= of X_in2] (d) {};
\draw[decoration={brace,raise=5pt},decorate]
  (a) -- node[right=6pt] {$X_\out$} (b);
\draw[decoration={brace,mirror,raise=5pt},decorate]
(c) -- node[left=6pt] {$X_\inp$} (d);
\end{tikzpicture}
\end{displaymath}
A morphism $f\colon X\to Y$ in this category is a pair of functions\Hair\footnote{In reality, these are not just arbitrary functions, rather generated by projections, diagonals and switchings; for more details consult Spivak \cite[Def.~3.3]{SpivakSteadyStates}. %
}
\begin{equation}\label{eq:wirdiag}
  \begin{cases}
  f_\inp\colon X_\out\times Y_\inp\to X_\inp \qquad\quad(a)\\
  f_\out\colon X_\out\to Y_\out\qquad\qquad\;\;(b)
  \end{cases}
\end{equation}
thought of as providing the flow of information in a picture as follows
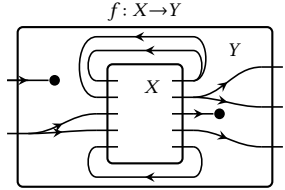
\begin{displaymath}%
\begin{tikzpicture}[oriented WD,baseline=(Y.center), bbx=2em, bby=1.2ex, bb port sep=1.2]
\node[bb={5}{5}] (X) {};
\node[bb={2}{3}, fit={($(X.north east)+(0.7,1.7)$) ($(X.south west)-(.7,.7)$)}] (Y) {};
\node [circle,minimum size=4pt, inner sep=0, fill] (dot1) at ($(Y_in1')+(.5,0)$) {};
\node [circle,minimum size=4pt, inner sep=0, fill] (dot2) at ($(X_out3)+(.5,0)$) {};
\draw[ar] (Y_in1) to (dot1);
\draw[ar] (X_out3) to (dot2);
\draw[ar] (Y_in2') to (X_in4);
\draw[ar] (Y_in2') to (X_in3);
\draw[ar] (X_out4) to (Y_out3');
\draw[ar] (X_out2) to (Y_out1');
\draw[ar] (X_out2) to (Y_out2');
\draw[ar] let \p1=(X.north west), \p2=(X.north east), \n1={\y1+\bby}, \n2=\bbportlen in
	(X_out1) to[in=0] (\x2+\n2,\n1) -- (\x1-\n2,\n1) to[out=180] (X_in1);
\draw[ar] let \p1=(X.north west), \p2=(X.north east), \n1={\y1+2*\bby}, \n2=\bbportlen in
	(X_out1) to[in=0] (\x2+\n2,\n1) -- (\x1-\n2,\n1) to[out=180] (X_in2);
\draw[ar] let \p1=(X.south west), \p2=(X.south east), \n1={\y1-\bby}, \n2=\bbportlen in
	(X_out5) to[in=0] (\x2+\n2,\n1) -- (\x1-\n2,\n1) to[out=180] (X_in5);	
\draw [label] node at ($(Y.north east)-(.5cm,.3cm)$) {$Y$}
node at ($(X.north east)-(.4cm,.3cm)$) {$X$}
node[above=of Y.north] {$f\colon X\to Y$}
;
\end{tikzpicture}
\end{displaymath}
which illustrates in diagrammatic view the system of equations~\ref{eq:wirdiag} (where the forks correspond to duplication and black bullets correspond to discarding). Information going through those wires can be anything insofar as the types match between ports.
The wires of the external input ports $Y_\inp$ can only go to the internal input ports $X_\inp$ (equation 1a), whereas the wires of the internal output ports $X_\out$ can be directed to the external output ports $Y_\out$ (equation 1b) as well as fed back to the internal input ports $X_\inp$ (equation 1a). 

This is a monoidal category, where the tensor product of any two labelled boxes $X$ and $Y$ is $X\otimes Y=(X_\inp\times Y_\inp, X_\out\times Y_\out)$ that represents the parallel placement of the two
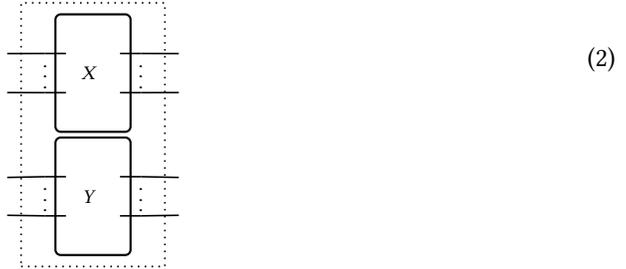
\begin{equation}\label{eq:tensorpic}
\begin{tikzpicture}[oriented WD,baseline=(Y.center), bbx=1.3em, bby=1ex, bb port sep=.12cm]
 \node[bb={2}{2}] (X1) {};
 \node[bb={2}{2},below =.5 of X1] (X2) {};
 \node[fit=(X1)(X2),draw,dotted] {};
 \draw[label] %
              node at ($(X1.west)+(1,0)$) {$X$}
              node at ($(X2.west)+(1,0)$) {$Y$}
              node at ($(X1.west)+(-.3,.4)$) {$\vdots$}
              node at ($(X1.east)+(.3,.4)$) {$\vdots$}
              node at ($(X2.west)+(-.3,.4)$) {$\vdots$}
              node at ($(X2.east)+(.3,.4)$) {$\vdots$};
 \draw (X1_in1) -- (-2.5,1.7);
 \draw (X1_out1) -- (2.5,1.7);
 \draw (X1_in2) -- (-2.5,-1.7);
 \draw (X1_out2) -- (2.5,-1.7);
  \draw (X2_in1) -- (-2.5,-9.1);
 \draw (X2_out1) -- (2.5,-9.1);
 \draw (X2_in2) -- (-2.5,-12.4);
 \draw (X2_out2) -- (2.5,-12.4);
 \end{tikzpicture}
\end{equation}
with input and output the (cartesian) product of the respective sets.
\end{definition}
For simplicity, we often abstract the pictures for objects, morphisms and tensor in $\ca{W}$ to 
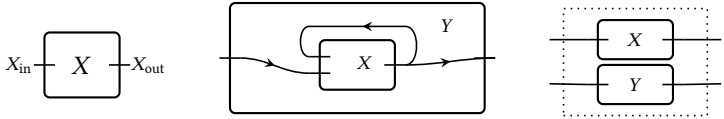
\begin{center}
\begin{tikzpicture}[oriented WD, bbx=.1cm, bby =.1cm, bb port sep=.15cm,baseline=(X.center)]
	\node [bb={1}{1}] (X) {$X$};
	\draw[label]
		node[left=.1 of X_in1]  {$X_\inp$}
        node[right=.1 of X_out1] {$X_\out$};
\end{tikzpicture}\qquad
 \begin{tikzpicture}[oriented WD,baseline=(X.center), bbx=2em, bby=1.2ex, bb port sep=1.2]
\node[bb={2}{1}] (X) {};
\node[bb={1}{1}, fit={($(X.north east)+(0.7,1.7)$) ($(X.south west)-(.7,.7)$)}] (Y) {};
\draw[ar] (Y_in1') to (X_in2);
\draw[ar] (X_out1) to (Y_out1);
\draw[ar] let \p1=(X.north west), \p2=(X.north east), \n1={\y1+\bby}, \n2=\bbportlen in
	(X_out1) to[in=0] (\x2+\n2,\n1) -- (\x1-\n2,\n1) to[out=180] (X_in1);
\draw [label] node at ($(Y.north east)-(.5cm,.3cm)$) {$Y$} node at ($(X.north east)-(.4cm,.3cm)$) {$X$}
node[above=of Y.north] {};
\end{tikzpicture}\qquad
\begin{tikzpicture}[oriented WD,baseline=(X2.north), bbx=1.3em, bby=1ex, bb port sep=.06cm]
 \node[bb={1}{1}] (X1) {$\scriptstyle X$};
 \node[bb={1}{1},below =.5 of X1] (X2) {$\scriptstyle Y$};
\node[fit=(X1)(X2),draw,dotted] {};
 \draw (X1_in1) -- (-2.5,0);
 \draw (X1_out1) -- (2.5,0);
  \draw (X2_in1) -- (-2.5,-3.85);
 \draw (X2_out1) -- (2.5,-3.85);
 \end{tikzpicture}
\end{center}
The composition in this category zooms two levels deep, and is formally defined as follows: for $f=(f_\inp,f_\out)\colon X\to Y$ and $g=(g_\inp,g_\out)\colon Y\to Z$ as in the systems of equations~\ref{eq:wirdiag}, the new wiring diagram $g\circ f\colon X\to Z$ consists of the functions
$\Big((g\circ f)_\inp\colon X_\out\times Z_\inp\to X_\inp,(g\circ f)_\out \colon X_\out\to Y_\out\Big)$
given by 
\begin{align*}
(g\circ f)_\inp(x',z)&=f_\inp(x',g_\inp(f_\out(x'),z)) \\
(g\circ f)_\out(x')&=g_\out(f_\out(x')).
\end{align*}
The identity morphism on $X$ is $(\pi_2\colon X_\out\times X_\inp\to X_\inp, 1_{X_\out}\colon X_\out\to X_\out)$, and the axioms of a category  hold.\Hair\footnote{There is a strong relation between $\ca{W}$ and the category of \emph{lenses}~\cite{BXtransformations}, as well as the \emph{Dialectica} category~\cite{DePaiva:1990}.} Moreover, the monoidal unit is the box $\SmallBox{\{*\}}{\{*\}}{I}$ and the axioms of a monoidal category can also be verified to hold~\cite{Vagner.Spivak.Lerman:2015a}.

The category $\ca{W}$ as defined above is really $\Set$-\emph{typed} or \emph{labeled}, namely the objects and morphisms are described using sets. 
However, the formalism allows to label the wires with any category equipped with finite products instead of $(\Set,\times,\{*\})$.
For example, the types could be in linear spaces $\mathbb{R}^n$ or topological spaces $(X,\tau)$ or even more general time-related categories like lists of signals expressed as sheaves on real-time intervals \cite[\S~3]{spivak:2016}. Not only do these different types accommodate systems with such inputs and outputs, but also often provide a passage between different models on the same system by functorially changing the types.

\begin{figure*}
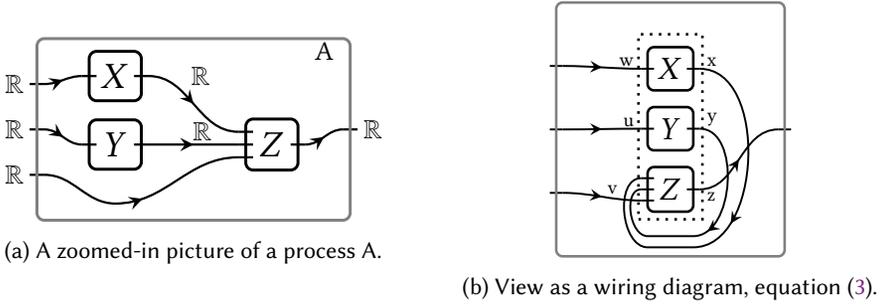

\begin{subfigure}{.45\textwidth}
  \centering
  \includestandalone[width=.8\linewidth]{./figures/minimal_example_correct}%
  \caption{A zoomed-in picture of a process A.}
  \label{min_ex}
\end{subfigure}
\begin{subfigure}{.45\textwidth}
  \centering
\includestandalone[width=.8\linewidth]{./figures/min_ex_tensor_step}
  \caption{View as a wiring diagram, equation (\ref{wirdiag}).}
  \label{min_ex_tensor}
\end{subfigure}
\caption{An example wiring diagram as a morphism in the category $\ca{W}$.}
\label{min_example}
\end{figure*}

The construction of this category allows us to formally give meaning to arbitrary wiring diagram pictures and as a result, coherently describe interconnections. As an example, consider three processes $X$, $Y$, and $Z$ (Fig.~\ref{min_ex}).
The involved labelled boxes are  $X=(\mathbb{R},\mathbb{R})$, $Y=(\mathbb{R},\mathbb{R})$ and $Z=(\mathbb{R}^3,\mathbb{R})$, which connected in the depicted way form the composite interface $A=(\mathbb{R}^3,\mathbb{R})$. Although $A$'s inputs and outputs are to the `outside world', they could also potentially interconnect to other boxes themselves.

To implement the above as a morphism in the category $\ca{W}$, we first `align' the boxes such that the wires follow their input and output (Fig.~\ref{min_ex_tensor}), which then forms a morphism from the tensor product of the three boxes $X\otimes Y\otimes Z$ (the dotted box) with input $\mathbb{R}^5$ and output $\mathbb{R}^3$, to the outside box $\mathrm{A}=(\mathbb{R}^3,\mathbb{R})$ with explicit description

\begin{equation}\label{wirdiag}
  \begin{cases}
  f_\inp\colon\overbrace{\mathbb{R}\times\mathbb{R}\times \mathbb{R}}^{(X\otimes Y\otimes Z)_\out}\times \overbrace{\mathbb{R}\times\mathbb{R}\times\mathbb{R}}^{A_\inp}\to\overbrace{\mathbb{R}\times\mathbb{R}\times\mathbb{R}\times\mathbb{R}\times\mathbb{R}}^{(X\otimes Y\otimes Z)_\inp}, & (x,y,z,w,u,v)\mapsto(w,u,x,y,v) \\
  f_\out\colon\underbrace{\mathbb{R}\times\mathbb{R}\times \mathbb{R}}_{(X\otimes Y\otimes Z)_\out}\to\underbrace{\mathbb{R}}_{A_\out}, & (x,y,z)\mapsto z
  \end{cases}.
\end{equation}
  
The two functions, $f_\inp$ and $f_\out$, specify 
which wires are connected to which; $f_\inp$ maps the three internal outputs $x,y,z$ together with the external inputs $w,u,v$ to the internal inputs, in the order determined by our alignment,\Hair\footnote{We could choose a different alignment of the internal boxes, which would result to a different, but essentially equivalent, pair of functions. This would not affect our analysis.} and $f_\out$ projects out of the three internal outputs $x,y,z$ the third one $z$. Even though Fig.~\ref{min_ex} does not involve any feedback loops on individual boxes, these are inherently used by the formalism as in Fig.~\ref{min_ex_tensor} in order to concretely form the wiring diagram equations (\ref{wirdiag}). 

To sum up, the category $\ca{W}$ provides a formal way of mathematically expressing any configuration at hand, with sole focus on the interconnection of vacant building blocks.

\section{Compositional Cyber-Physical Systems Theory}

Assessing the correct behavior of \textsc{cps} requires several model views. Before discussing them, we must first clarify the meaning of the terminology that we will use. We choose to use the terminology of \emph{requirements}, \emph{system behavior}, and \emph{system architecture} to describe the different diagrammatic abstractions of \textsc{cps} models. We define requirements as constraints over system behavior
and system architecture. By system behavior we mean models 
of the form of automata or state space models. 
By system architecture we mean models of candidate implementations
that, in the case of \textsc{cps}, include hardware and software
for the embedded system portion of \textsc{cps} and motors, control surfaces,
and mechanical structure for the physical portion of the \textsc{cps}. 
In the following formalism in general, we will view the individual diagram pictures
as architecture, and the particular semantics that go into the boxes
within this diagram as behavior, omitting the leading word `system' when only discussing about the diagrammatic representation.
Contracts that constraint both behavior and architecture
in this sense will represent a subset of system safety requirements.

The categorical approach has the advantage of providing a \emph{compositional} modeling and analysis, 
in which the composite system is completely and uniquely determined 
from its subsystems and their interconnections. This is achieved through the implementation of the formalism in two parts. The first is a behavior algebra that allows the hierarchical modeling between the abstraction of system behavior and system architecture, in a \emph{zoom-in, zoom-out} approach~\cite{willems:2007}, where each view may have distinct inputs and outputs.
The second is a contract algebra that applies constraints as defined in requirements over the behavior algebra.

An analogously high-level approach using monoidal categories and compositional techniques
has already found success in categorical quantum mechanics~\cite{abramsky:2009,coecke:2010}, where it has become the de facto language
to describe and manipulate quantum processes diagrammatically.
We posit that a similar innovation should take place
in the design and assessment of safety-critical \textsc{cps}, due to the concerns raised by 
the intertwined nature of digital control
with physical processes and the environment.
We will view distinct but related system models,
pertinent to assuring the correct behavior of \textsc{cps}, as \emph{algebras} 
of the monoidal category of wiring diagrams.

The wiring diagram approach diverges from input-output models.
While the diagrammatic syntax looks similar to such models,
what is contained within the boxes need not be a mathematical function.
It can instead be any sort of process, from very concrete descriptions like automata, to more abstract processes which could be deterministic or non-deterministc, to mere requirements of a mathematically unknown formula. 
Similarly, the arrows do not need to contain one piece of information,
for example the input and output of a function;
rather, arrows can carry arbitrary objects of a chosen category of types.
 Previous compositional modeling methods for \textsc{cps} are often limited to sets and functions or in the most general sense, relations.  However, the state space of a controls system need not be the set $\mathbb{R}$, but could instead be a topological space like the line or circle $\mathbb{S}$. %
 The rich interplay between topology and category theory
 positions category theory as a particularly good candidate
 for modeling dynamics, for example see Hansen and Ghrist~\cite{hansen:2020}
 or earlier, in the more related area of hybrid systems, Ames~\cite{ames:2006} and Tabuada et al.~\cite{tabuada:2002}.

We now develop the formalism for the three system views necessary to assess the correct behavior of \textsc{cps}: system behavior, system architecture, and (a subset of) requirements.

\subsection{System Behavior via Algebras on the Category \texorpdfstring{$\ca{W}$}{\textbf{W}} }\label{sec:systemsasalgebras}

The category of wiring diagrams does not populate the boxes with actual systems, for example, dynamical systems (\cref{sec:categoryW}).
This is instead done by developing extra structure on top of it. By knowing the configuration of the component systems, the composite system can then be uniquely determined.

Categorically, this is described as an \emph{algebra} on $\ca{W}$, namely a lax monoidal functor $F\colon(\ca{W},\otimes, I)\to(\mathbf{Cat},\times,\mathbf{1})$. The idea is that each algebra assigns to a box $X=(X_\inp,X_\out)$ a category $FX$ of systems that can be placed in the box, and also assigns to a wiring diagram $f=(f_\inp,f_\out)$ a functor $Ff\colon FX\to FY$ that, given a system $s$ inhabiting the internal box of a wiring diagram, produces the \emph{composite system} $F(f)(s)$ inhabiting the external box. %
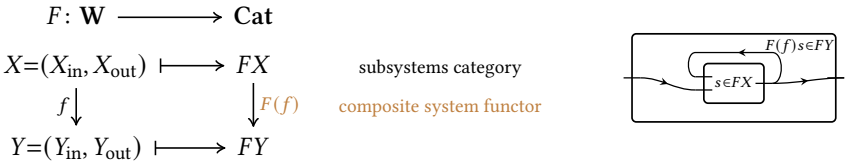
\begin{displaymath}
 \begin{tikzcd}[row sep=.05in]
F\colon\ca{W}\ar[r] & \Cat &\\
X{=}(X_\inp,X_\out) \ar[r,mapsto]\ar[dddd,"f"'] & 
FX\ar[dddd,"{\color{brown}F(f)}"] & \scriptstyle\textrm{subsystems category}
\ar[dddd,phantom,"{\scriptstyle\color{brown}\textrm{composite system 
functor}}"] \\
&&& \\
&&& \\
&&& \\
Y{=}(Y_\inp,Y_\out) \ar[r,mapsto] & FY & \phantom{ABC}
 \end{tikzcd}\quad
 \scalebox{.8}{\begin{tikzpicture}[oriented WD,baseline=(X.center), bbx=2em, bby=1.2ex, bb port sep=1.2]
\node[bb={2}{1}] (X) {};
\node[bb={1}{1}, fit={($(X.north east)+(0.7,1.7)$) ($(X.south west)-(.7,.7)$)}] (Y) {};
\draw[ar] (Y_in1') to (X_in2);
\draw[ar] (X_out1) to (Y_out1);
\draw[ar] let \p1=(X.north west), \p2=(X.north east), \n1={\y1+\bby}, \n2=\bbportlen in
	(X_out1) to[in=0] (\x2+\n2,\n1) -- (\x1-\n2,\n1) to[out=180] (X_in1);
\draw [label] node at ($(Y.north east)-(.6cm,.2cm)$) {$F(f)s\in FY$} node at ($(X.north east)-(.5cm,.3cm)$) {$s\in FX$}
node[above=of Y.north] {};
\end{tikzpicture}}
\end{displaymath}
Intuitively, the object assignment 
$FX$ and $FY$ gives semantics 
to arbitrary boxes through the subsystems category 
while the composite system functor $Ff$ 
assembles the composite operations 
of the overall system behavior.
Moreover, the monoidal structure of the functor via the laxator $\phi_{X,Y}\colon FX\times FY\to F(X\otimes Y)$ ensures that for given systems inside parallely placed boxes, we can always determine a system inhabiting their tensor product
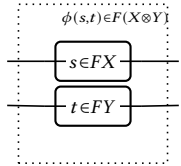
\begin{displaymath}
\begin{tikzpicture}[oriented WD,baseline=(X2.north), bbx=1.3em, bby=1ex, bb port sep=.06cm]
 \node[bb={1}{1}] (X1) {$\scriptstyle s\in FX$};
 \node[bb={1}{1},below =.5 of X1] (X2) {$\scriptstyle t\in FY$};
\node[fit=(X1)(X2),draw,dotted,inner sep=14pt] (Y) {};
 \draw (X1_in1) -- (-2.5,0);
 \draw (X1_out1) -- (2.5,0);
  \draw (X2_in1) -- (-2.5,-3.85);
 \draw (X2_out1) -- (2.5,-3.85);
 \draw [label] node at ($(Y.north east)-(.7cm,.2cm)$) {$\scriptscriptstyle\phi(s,t)\in F(X\otimes Y)$};
 \end{tikzpicture}
\end{displaymath}
The categorical formulation allows us
to use a number of algebras according to our purposes. Below we describe two such algebras of discrete dynamical systems, and later we will examine the algebra of contracts (Section~\ref{sec:contracts}). There exist also other algebras, describing systems behaviors that are not like difference equations. For example, algebras for abstract total or deterministic machines~\cite{spivak:2016}.

The diagrammatic representation via wiring diagrams
for system modeling and analysis is rather straightforward,
particularly because wiring diagrams are similar
to engineering block diagrams
and, hence, the visual syntax is equivalent to existing \textsc{cps} design tools. However, the current diagrammatic representation is mathematically richer and more concrete -- it also accounts for actual composition computations as we will see below.
Another important factor specifically for \textsc{cps} is the richness
of other possible algebras or semantics that one can develop and assign
in these boxes using as backing the notion of the monoidal category.
As \textsc{cps} become more complex, cooperative, and coordinated
these functorial semantics can give formal relations
between several concepts important in modeling and assurance
of safe \textsc{cps}~\cite{cps}.

\subsubsection{Moore Machines}

As an illustrative example on how to develop and use the behavior algebra on an architecture in $\ca{W}$, we will position the familiar Moore machines 
inside the boxes $X$, $Y$ and $Z$ of Fig.~\ref{min_ex}. This is a simple yet useful demonstration of the algebra machinery because Moore machines model discrete dynamical systems.
To concretely describe the systems composite, we first need to verify that Moore machines form a $\ca{W}$-algebra. Indeed, there is a monoidal functor 
\begin{displaymath}
 \mathcal{M}\colon\ca{W}\to\Cat
\end{displaymath}
which maps each $(X_\inp,X_\out)$ to the category $\mathcal{M}(X_\inp,X_\out)$ where
\begin{itemize}
    \item objects are triples $(S,u,r)$ where $S$ is the \emph{state space} set, $u\colon S\times X_\inp\to S$ is the \emph{update function} and $r\colon S\to X_\out$ is the \emph{readout function};
    \item morphisms $(S,u,r)\to(S',u',r')$ are functions $f\colon S\to S'$ between the state spaces that commute with the update and readout functions, namely $f(u(s,x))=u'(fs,x)$ and $f(r(s))=r'(fs)$.
\end{itemize}
Hence, $\mathcal{M}(X_\inp,X_\out)$ is the category of Moore machines with fixed input and output alphabet $X_\inp$ and $X_\out$ respectively. For example, an object of the category $M(\{0,1\},\{0,1\})$ with inputs and outputs the booleans, is the `\texttt{not}' finite state machine
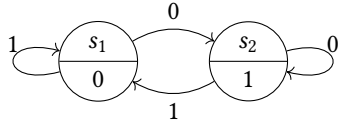
\begin{displaymath}
\begin{tikzpicture}
  \node[circle split, draw] (s1) at (0,0) {$s_1$ \nodepart{lower} 0};
  \node[circle split, draw] (s2) at (2,0) {$s_2$ \nodepart{lower} 1};
  \draw[->] (s1) to [bend left] node[above] {0} (s2);
  \draw[->] (s2) to [bend left] node[below] {1} (s1);
  \draw[->] (s1) to [loop left] node[above] {1} (s1);
  \draw[->] (s2) to [loop right] node[above] {0} (s2);
\end{tikzpicture}
\end{displaymath}
with state space $S=\{s_1,s_2\}$ and update and readout functions depicted in the above state diagram, for example, $u(s_1,0)=s_2$ (middle top edge) and $r(s_2)=1$ (bottom part of $s_2$-node). 

Having defined the categories of systems that can inhabit boxes in wiring diagram pictures for this specific Moore machine model, we proceed to define the composite system functor $\mathcal{M}(f)\colon \mathcal{M}X\to \mathcal{M}Y$, given a wiring diagram $f=(f_\inp,f_\out)\colon X\to Y$. Explicitly, this functor maps a Moore machine $(S,u,r)$ with input and output $X_\inp,X_\out$ to a Moore machine $(S,u',r')$ with input and output $Y_\inp,Y_\out$ having the \emph{same} state space $S$, but with new update and readout functions formed as follows
\begin{align}
&u'\colon Y_\inp\times S\to S, & u'(y,s)&=u(f_\inp(y,r(s)),s) \label{eq:Mf}\\
&r'\colon S\to Y_\out, & r'(s)&=f_\out(r(s)) \nonumber
\end{align}
Finally, we need to specify the monoidal structure of $\mathcal{M}$ by providing functors $\mathcal{M}(X)\times \mathcal{M}(Y)\to \mathcal{M}(X\otimes Y)$. Explicitly, given two Moore machines $(S_X,u_X\colon X_\inp\times S_X\to S_X,r_X\colon S_X\to X_\out)$ and $(S_Y,u_Y\colon Y_\inp\times S_Y\to S_Y,r_Y\colon S_Y\to Y_\out)$, we construct a new Moore machine with space set $S_X\times S_Y$ and update and readout functions
\begin{gather}
u\colon X_\inp\times Y_\inp \times S_X\times S_Y\to S_X\times S_Y, \qquad u(x,y,s,t)=(u_X(x,s),u_Y(y,t)) \label{eq:laxator}\\
r\colon S_X\times S_Y\to X_\out\times Y_\out, \qquad r(s,t)=(r_X(s),r_Y(t)) \nonumber
\end{gather}

It can been be verified that with the above assignments, Moore machines satisfy the axioms of a wiring diagram algebra \cite[\S 2.3]{spivak:2016}. We can therefore arbitrarily interconnect such systems, in particular as in Fig.~\ref{min_ex}, and produce a new such system with a description only in terms of its components and their wiring. Suppose we have Moore machines in the boxes $\SmallBox{}{}{X}$, $\SmallBox{}{}{Y}$, $\SmallBoxThree{}{}{}{}{Z}$, all with $\mathbb{R}$-valued wires, with state spaces $S_X$, $S_Y$ and $S_Z$ and update and readout functions respectively as in
\begin{displaymath}
\begin{cases}
S_X\times\mathbb{R}\xrightarrow{u_X} S_X \\
S_X\xrightarrow{r_X}\mathbb{R}
\end{cases}
\begin{cases}
S_Y\times\mathbb{R}\xrightarrow{u_Y} S_Y \\
S_Y\xrightarrow{r_Y}\mathbb{R}
\end{cases}
\begin{cases}
S_Z\times\mathbb{R}^3\xrightarrow{u_Z} S_Z \\
S_Z\xrightarrow{r_Z}\mathbb{R}.
\end{cases}
\end{displaymath}
The algebra machinery \cref{eq:Mf,eq:laxator} for the specific wiring diagram \cref{wirdiag} produces the composite Moore machine which inhabits the outer box $\SmallBoxThree{}{}{}{}{A}$
with state space $S_X\times S_Y\times S_Z$, readout function $r\colon S_X\times S_Y\times S_Z\to \mathbb{R}$ given by $(s,t,p)\mapsto r_Z(p)$ and update function $S_X\times S_Y\times S_Z\times\mathbb{R}^3\to S_X\times S_Y\times S_Z$ given by  
\begin{displaymath}
(s,t,p,w,u,v)\mapsto \left(u_X(s,w),
u_Y(t,u),u_Z(p,r_X(s),r_Y(t),v)\right).
\end{displaymath}

In general, the composite system is produced using the algebra machinery, no matter how complicated the systems or the wiring diagram is: given any type-respecting interconnection involving arbitrary feedback loops or parallel/serial arrangements, the monoidal functor will determine a result. Therefore, this functoriality alleviates some of the scalability issues present in other formalisms. As we will see later, often some pre-existing knowledge on the desired behavior of a composite system can possibly inform not only the components' behavior but also the choice of wiring.

\subsubsection{Linear Time-Invariant Systems}\label{sec:LTIS}

There is a sub-algebra of the algebra of Moore machines, for \emph{linear time-invariant systems} (\textsc{ltis}) or linear discrete dynamical systems per Spivak~\cite{SpivakSteadyStates}. In fact, the Moore machines model is an algebra of $\ca{W}_\Set$, where the types of wires are sets and the wiring diagrams are given by functions, whereas the \textsc{ltis} model is an algebra of $\ca{W}_\Lin$, where the types are given by $\Lin$, the category of linear spaces and linear maps.

Explicitly, there is a monoidal functor $\mathcal{L}\colon \ca{W}_\Lin\to\Cat$ that assigns to any box $\SmallBox{X_\inp}{X_\out}{}$ a category $\mathcal{L}(X_\inp,X_\out)$ of systems $(S,u\colon S\times X_\inp\to S,r\colon S\to X_\out)$ like before, but where all $S,X_\inp,X_\out$ are linear spaces and both update and readout functions $u$ and $r$ are linear functions expressed as
\begin{align*}
u(s,x)=\;\; & \mathscr{A}\cdot s+ \mathscr{B}\cdot x=\begin{pmatrix}\mathscr{A} & \mathscr{B}
\end{pmatrix}
\begin{pmatrix}
s \\
x
\end{pmatrix}
\\
r(s)=\;\; & \mathscr{C}\cdot s
\end{align*}
where $\mathscr{A}$, $\mathscr{B}$ and $\mathscr{C}$ are matrices of appropriate dimension. For example, if the input, output and state spaces are $X_\inp=\RR^k$, $X_\out=\RR^\ell$ and $S=\RR^n$, then
\begin{equation}\label{eq:dimensions}
\begin{cases}
\mathscr{A}\in {}_nM_n & \textrm{represents a linear transformation } \RR^n\to\RR^n \\
\mathscr{B}\in {}_nM_k & \textrm{represents a linear transformation }  \RR^k\to\RR^n \\
\mathscr{C}\in {}_\ell M_n & \textrm{represents a linear transformation } \RR^n\to\RR^\ell.
\end{cases}
\end{equation}

Now given an arbitrary wiring diagram $f=(f_\inp, f_\out)\colon (X_\inp,X_\out)\to (Y_\inp,Y_\out)$ as formalized in the system of equations~\cref{eq:wirdiag}, where for $Y_\inp=\RR^{k'}$ and $Y_\out=\RR^{\ell'}$ both linear functions of the wiring diagram are also expressed as corresponding matrices $f_\inp=\begin{pmatrix}
{}_k(\mathscr{A}^f)_{\ell} & {}_k(\mathscr{B}^f)_{k'}
\end{pmatrix}$
and $f_\out={}_{\ell'}\mathscr{C}^f_{\ell}$, the functor $\mathcal{L}(f)$ maps some system $(S,\mathscr{A},\mathscr{B},\mathscr{C})$ in $\SmallBox{\RR^k}{\RR^\ell}{X}$ to the system 
\begin{equation}\label{eq:LTIScomposite}
(S,\mathscr{A}+\mathscr{B}\cdot \mathscr{A}^f\cdot \mathscr{C},\; \mathscr{B}\cdot \mathscr{B}^f,\; \mathscr{C}^f\cdot \mathscr{C})
\end{equation}
in $\SmallBox{\RR^{k'}}{\RR^{\ell'}}{Y}$. The earlier-used term \emph{sub-algebra} precisely means that this formula is a special case of equation \cref{eq:Mf} when the functions involved are of this specific form.

Finally, the monoidal structure of this assignment $\mathcal{L}\colon\mathbf{W}_\Lin\to\Cat$ is given by functors $\mathcal{L}(X)\times \mathcal{L}(Y)\to \mathcal{L}(X\otimes Y)$ that map any two such systems $(S_X,\mathscr{A}_X,\mathscr{B}_X,\mathscr{C}_X)$ and $(S_Y,\mathscr{A}_Y,\mathscr{B}_Y,\mathscr{C}_Y)$ inhabiting parallel boxes as in wiring diagram \cref{eq:tensorpic} give rise to a parallel composite system 
\begin{displaymath}
\left(S_X\times S_Y,\begin{pmatrix}
\mathscr{A}_X & 0 \\
0 & \mathscr{A}_Y
\end{pmatrix},
\begin{pmatrix}
\mathscr{B}_X & 0 \\
0 & \mathscr{B}_Y
\end{pmatrix},
\begin{pmatrix}
\mathscr{C}_X & 0 \\
0 & \mathscr{C}_Y
\end{pmatrix}\right).
\end{displaymath}

\subsubsection{Functions (as a Non-Example)}\label{sec:functions}

If we would like to populate the boxes of a wiring interconnection with mathematical functions, namely assign to some $\SmallBox{X_\inp}{X_\out}{X}$ a function $h\colon X_\inp\to X_\out$, there is no natural way to make this assignment into an algebra $\ca{W}\to\Cat$. The main reason this fails is the existence of the feedback loop.

However, we can incorporate functions into other existing models, for example Moore machines. It is possible to express a function $h\colon X_\inp\to X_\out$ as an object of $\mathcal{M}(X_\inp,X_\out)$, with state space the domain $X_\inp$ and update and readout functions $\pi_2\colon X_\inp\times X_\inp\to X_\inp$ projecting the second variable and $h\colon X_\inp\to X_\out$ applying the said function. The resulting finite state machine at each round replaces the old input with the new input, and outputs the function application on it. Analogously, a linear function can be viewed as a linear time-invariant system if we set $\mathscr{A}=0$ the zero matrix, $\mathscr{B}=I$ the unit matrix and $\mathscr{C}=h$ the matrix represents the given linear transformation. 

As a result, functions can be indeed used to populate boxes, and wired with other functions or Moore machines they produce a composite Moore machine using the algebra $\mathcal{M}\colon\ca{W}\to\Cat$. It is also the case that sometimes, wiring two functions using the Moore machine algebra machinery, we end up with another function and not a more general Moore machine -- this usually happens in serial-like wirings without loops.

Summarizing this section, the starting point is the category of wiring diagrams $\ca{W}$ with no processes inside the boxes.
We can then assign the behavior of Moore machines inside the boxes using the corresponding $\ca{W}$-algebra $\mathcal{M}$, or the behavior of linear discrete dynamical systems using the sub-algebra of linear time-invariant systems $\mathcal{L}$,
which recovers the standard model of state-space representation in modern control
albeit with a slightly different syntax.
By composing behaviors using the latter, we recover a block-diagonal state space model,
a useful representation for modeling the control portion of \textsc{cps}. 
 
To model state-space representations we had to develop all the above categorical machinery.
However, the point of modeling algebraically is
that now we can ensure composition
and also caution when two boxes do not compose
in the strict mathematical sense
using only the diagrammatic syntax,
which is familiar to develop and manipulate.
At the moment, we have illustrated a couple of examples
of behavioral models that can inhabit the boxes,
but the algebraic machinery is not limited by those.
We could, for example, inhabit the boxes
with hybrid systems, while at the same time composing properly -- whose detailed illustration is part of future work where time is also incorporated in the framework. Here, in the absence of the time element, behaviors are to be observed in an instantaneous way, for example, feedback loops do not produce delay effects. Such issues shall be tackled when the wires carry time-sensitive data.

At the moment we have developed the theory using pen and paper,
but it is possible to produce an algorithmic implementation
of this algebra machinery and then enforce these rules diagrammatically.
Particularly, recent software implementations of categorical concepts and specifically symmetric monoidal categories and wiring diagrams into software can be used to develop prototype tools and analyses using compositional \textsc{cps} theory~\cite{patterson:2021}.
This is useful because diagrammatic languages do not scale with system design complexity~\cite{myers:1990} and this computational implementation can allow us to both visualize important subsystems but also implement these categories and algebras using a programming language in the same way Modelica, Lustre, or Lingua Franca~\cite{lohstroh:2020}.

\subsection{System Architecture via Hierarchical Decomposition}
\label{sec:theory:arch}

Starting with a \textsc{cps} from a designer point of view, we now might want to model a candidate system architecture. In general, decomposing a \textsc{cps} in certain sub-components and using a specific wiring between them follows some choices based on the physical reality, experience, purpose and access to particular components at the time.
Having formalized an agnostic process interface where various descriptions could live on as an object in the category of wiring diagrams $\ca{W}$, as well as arbitrary zoomed-in pictures of a system as a morphism in $\ca{W}$, we have now access to all necessary tools to realize the above system architecture design process using the general notion of a \emph{slice category}.

For any category $\ca{C}$ and a fixed object $C\in\ca{C}$, the slice category $\ca{C}/C$ has as objects $\ca{C}$-morphisms with fixed target $C$, for example $f\colon A\to C, g\colon B\to C,\cdots$. The arrows in that category from some $f$ to some $g$ are $\ca{C}$-morphisms $k\colon A\to B$ between the domains, making the formed triangle
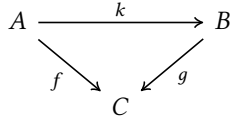
\begin{displaymath}
\begin{tikzcd}
A\ar[rr,"k"]\ar[dr,"f"'] && B\ar[dl,"g"] \\
& C &
\end{tikzcd}
\end{displaymath}
commute, namely $g\circ k=f$. This data forms a category, which also illustrates the abstract nature of the initial category definition (\cref{sec:basiccats}): objects and arrows can be of any sort (in this case objects are morphisms of a certain shape in some fixed category, and arrows are also morphisms that satisfy a property) as long as they satisfy the axioms of a category.

For our wiring diagram category $\ca{W}$, where a morphism $f\colon X\to Y$ can be thought of as an implementation of an interface $Y$ into sub-interface(s) $X$ wired in a specific manner, the slice category $\ca{W}/Y$ of all arrows mapping into the chosen object $Y$ essentially contains all possible design choices available to a system engineer.
This formally captures the possibility of implementing a system in multitudes of ways.

Concretely, suppose we have a system with $\mathbb{R}^3$-inputs and $\mathbb{R}$-outputs, namely inhabiting a box $\SmallBoxThree{\mathbb{R}}{\mathbb{R}}{\mathbb{R}}{\mathbb{R}}{A}$. How can we decompose it into sub-processes, and how can they be interconnected to form the given system? All the possible decompositions can thus be thought of as the objects of the slice category $\ca{W}/A$. For example, Fig.~\ref{min_ex} depicts one of these choices, namely the specific wiring diagram $f\colon X\otimes Y\otimes Z\to A$. 

\begin{figure}
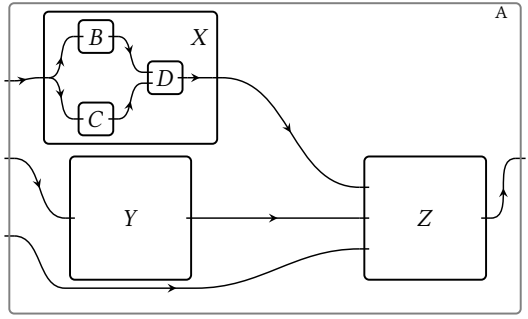

  \centering
  \includestandalone[width=.5\linewidth]{./figures/zoomin2}%
  \caption{A two-level zoomed-in picture of a process A.}
 \label{zoomin2}
\end{figure}

Now suppose we make another implementation choice to further decompose the box $X$ as in
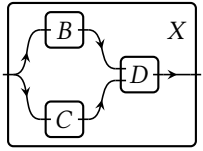
\begin{displaymath}
\begin{tikzpicture}[oriented WD, bbx = .5cm, bby =.8ex, bb min width=.5cm, bb port length=2pt, bb port sep=1,baseline=.5]
\node[bb={2}{1}] (P) {$D$};
\node[bb={1}{1},above left=of P] (I) {$B$};
\node[bb={1}{1},below left=of P] (J) {$C$};
\node[bb={1}{1},inner sep=.1in,fit={(J)(I)(P)}] (L) {};
\draw[ar] (L_in1') to (I_in1);
\draw[ar] (L_in1') to (J_in1);
\draw[ar] (I_out1) to (P_in1);
\draw[ar] (J_out1) to (P_in2);
\draw[ar] (P_out1) to (L_out1');
\node[right=1cm of I]{$X$};
 \end{tikzpicture}
\end{displaymath}
meaning we choose a specific wiring diagram $g\colon B\otimes C\otimes D\to X$. This constitutes another level of zoom-in for the process in $A$, at least for the subcomponent $X$ (Fig.~\ref{zoomin2}). Categorically, this is a picture of the composite morphism $f\circ(g\otimes1\otimes1)$, represented with the dashed arrow below
\begin{displaymath}
\begin{tikzcd}
(B\otimes C\otimes D)\otimes Y\otimes Z\ar[rr,"g\otimes1\otimes1"]\ar[dr,dashed] && X\otimes Y\otimes Z\ar[dl,"f"] \\
& A &
\end{tikzcd}
\end{displaymath}
where the top arrow employs the morphism $g$ as the implementation of $X$ and identity morphisms on $Y$ and $Z$ (as trivial implementations), and $f$ is the earlier $A$-implementation  (Fig.~\ref{min_ex}). In the end, we can disregard the borders of the interface $X$ and map directly from the subcomponents $B\otimes C\otimes D\otimes Y\otimes Z$ to $A$ without passing through $X$ at all if desired. 
As a result, we are free to use hierarchical decomposition of processes for any sub-component (or for many simultaneously) and each time, these architectural choices add one more composite morphism to the resulting wiring diagram that expresses an implementation of the outmost system process.

\subsection{System Requirements via Contracts}
\label{sec:contracts}

The concept of a \emph{contract}, fundamental for this work, is another example of an algebra for the monoidal category of labeled boxes and wiring diagrams $\ca{W}$.
 In detail, for any labeled box $X=(X_\inp,X_\out)$, a contract is defined to be a relation $$R\subseteq X_\inp\times X_\out$$ expressing the \emph{allowable} tuples of input and output behaviors of the process. Such a description is one among the most widespread abstract systems modeling notions, see for example Mesarovic and Takahara~\cite[\S 2]{Mesarovic:1989}. We make a distinction between the explicit defining process of a system; that is, the behavior assigned to a wiring diagram, and the system behavior. However, abstractly a system \emph{is} its behavior and therefore modeling a system in the wiring diagram paradigm makes those two notions equivalent. The distinction is however useful for separating the \emph{behavior} algebra from the \emph{contracts} algebra, which are formally related but can be used independently of each other.

\subsubsection{Static Contracts}\label{sec:staticcontracts}

The algebra of \emph{static contracts} is a variation of the algebra originally developed by Schultz et al.~\cite[\S 4.5]{spivak:2016}. Explicitly, the functor $\mathcal{C}\colon\ca{W}\to\Cat$ bound to express conditions on inputs and outputs in a time-less manner, assigns to a box $\SmallBox{X_\inp}{X_\out}{X}$ the category $\mathcal{C}(X_\inp,X_\out)$ of binary relations, that is, subsets $i\colon R\hookrightarrow X_\inp\times X_\out$, with morphisms $f\colon R\to P$ being subset inclusions of the form
\begin{displaymath}
\begin{tikzcd}
R\ar[r,hook]\ar[d,hook,"f"'] & X_\inp\times X_\out \\
P\ar[ur, hook] &
\end{tikzcd}
\end{displaymath}
For a given contract $R_X\subseteq X_\inp\times X_\out$ and a wiring diagram $(f_\inp\colon X_\out\times Y_\inp\to X_\inp, f_\out\colon X_\out\to Y_\out)$, the application of the functor $\mathcal{C}(f)$ on $R_X$ is the contract $R_Y\subseteq Y_\inp\times Y_\out$ described by
\begin{equation}\label{eq:compositecontract}
R_Y=\{(y_1,y_2)\in Y_\inp\times Y_\out\;|\;\exists x_2\in X_\out \textrm{ such that }(f_\inp(x_2,y_1),x_2)\in R_X\textrm{ and }f_\out(x_2)=y_2\}.
\end{equation}
This formula arises categorically (\cref{sec:appendix}).
In various examples, this composite contract may be expressed in more elementary terms depending on the form of the component contracts $R_X$ and the wiring diagram at hand. 

For the monoidal structure of the functor, suppose we have two parallel boxes \cref{eq:tensorpic} with contracts $R_X\subseteq X_\inp\times X_\out$ and $R_Y\subseteq Y_\inp\times Y_\out$. The laxator $\phi_{X,Y}\colon \mathcal{C}(X)\times \mathcal{C}(Y)\to \mathcal{C}(X\otimes Y)$ induces a contract on the box $(X_\inp\times Y_\inp,X_\out\times Y_\out)$ which is merely the cartesian product
\begin{displaymath}
R_X\times R_Y\hookrightarrow X_\inp\times X_\out\times Y_\inp\times Y_\out\xrightarrow{\cong}X_\inp\times Y_\inp\times X_\out\times Y_\out
\end{displaymath}
that essentially switches the two middle variables, $$\phi_{X,Y}(R_X,R_Y)=\{(x_1,y_1,x_2,y_2)\;|\;(x_1,x_2)\in R_X\textrm{ and }(y_1,y_2)\in R_Y\}.$$

As an example, suppose we ask that some process in $X$ (Fig.~\ref{min_ex}) satisfies the contract $R_X\subseteq\mathbb{R}\times\mathbb{R}$, some process in $Y$ satisfies the contract $R_Y\subseteq\mathbb{R}\times\mathbb{R}$ and some process in $Z$ satisfies the contract $R_Z\subseteq\mathbb{R}^3\times\mathbb{R}$. The fact that contracts form an algebra on $\ca{W}$ ensures that the composite process in $A$ will necessarily satisfy a contract formed only in terms of $R_X, R_Y$ and $R_Z$ and their interconnection $(f_\inp,f_\out)$, and specifically
\begin{displaymath}
R_A=\{(w,u,v,z)\in\mathbb{R}^4\;|\;\exists (x,y)\in\mathbb{R}^2 \textrm{ such that }
(w,x)\in R_X,(u,y)\in R_Y, (x,y,v,z)\in R_Z\}.
\end{displaymath}
The algebra machinery produces a contract that matches our intuition: whenever the interconnected composite in Fig.~\ref{min_ex} receives three real numbers $(w,u,v)$ as inputs, it must produce an output $z$ which is $R_Z$-allowable by (i.e. related to) $(x,y,v)$, for some real $x$ which is $R_X$-allowable by $w$ and some real $y$ which $R_Y$-allowable by $u$. Not all inputs of this composite $A$ will have an allowable output, and that completely depends on the contracts of its components $X$, $Y$ and $Z$. 

As another example, which highlights the strong connection between the contract algebra machinery and the usual relation operators, consider a simple wiring diagram with $\mathbb{R}$-typed wires on the left, expressing serial composition of two boxes: 
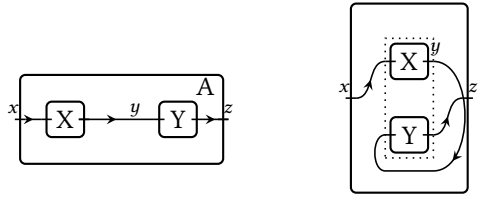
\begin{displaymath}%
\begin{tikzpicture}[oriented WD, bbx = .5cm, bby =.8ex, bb min width=.5cm, bb port length=2pt, bb port sep=1]
\node[bb={1}{1}] (X) {X};
\node[bb={1}{1},right=1cm of X] (Y) {Y};
\node[below=.5cm of X] () {};
\node[bb={1}{1},fit=(X)(Y),inner sep=10pt] (A) {};
\draw[ar] (X_out1') to (Y_in1);
\draw[ar] (A_in1') to (X_in1);
\draw[ar] (Y_out1) to (A_out1);
\draw[label] node [below left=.3 of A.north east] {A}
 node [above right=.2 and .2 in of X_out1] {$\scriptstyle y$}
node [above left= of A_in1] {$\scriptstyle x$}
node [above right= of A_out1] {$\scriptstyle z$}
;
\end{tikzpicture}\qquad\qquad
\begin{tikzpicture}[oriented WD, bbx = .5cm, bby =.8ex, bb min width=.5cm, bb port length=2pt, bb port sep=1]
\node[bb={1}{1}] (X) {X};
\node[bb={1}{1},below=.5cm of X] (Y) {Y};
\node[bb={1}{1},fit=(X)(Y),inner sep=15pt] (A) {};
\node[fit=(X)(Y),inner sep=2pt,dotted,draw] (A) {};
 \draw[ar] let \p1=(Y.south east), \p2=(Y.south west), \n1={\y1-2*\bby},\n2=\bbportlen in (X_out1) to [in=0] (\x1+\n2,\n1) -- (\x2-\n2,\n1) to [out=180] (Y_in1);
\draw[ar] (A_in1') to (X_in1);
\draw[ar] (Y_out1) to (A_out1');
\draw[label]
node [above right= of X_out1] {$\scriptstyle y$}
node [above left= of A_in1] {$\scriptstyle x$}
node [above right= of A_out1] {$\scriptstyle z$};
\end{tikzpicture}
\end{displaymath}
This morphism $f\colon X\otimes Y\to A$ is described by $(f_\inp(y,z,x)=(x,y),f_\out(y,z)=z)$ according to its equivalent arrangement on the right, and given two contracts $R_X$ and $R_Y$ the formula \cref{eq:compositecontract} produces the composite contract
$$R_A=\{(x,z)\;|\;\exists y\textrm{ such that }(x,y)\in R_X\textrm{ and }(y,z)\in R_Y\}$$
which is the usual composition of binary relations.

What is particularly interesting about this algebra of contracts is that it is `agnostic' to the exact specification of the systems. This means that although categorically it is expressed the same way as, for example, Moore machines, it is of a quite different flavor: we are not interested in giving explicit functions that describe the composite process, but in expressing all the possible (input,output) pairs that can be observed on it. This is very convenient especially when connecting systems of different models, for example, a Moore machine with an `abstract machine' \cite[\S 4]{spivak:2016}. Even if we cannot compose them in the previous sense, since they form distinct algebras (that is,  they are described by different functors $\ca{W}\to\mathbf{Cat}$), we can still compose and examine the requirements the composite satisfies, in this relational sense.

\subsubsection{Independent Contracts}\label{sec:independentcontr}

We will also be interested in a subclass of static contracts, called \emph{independent}, of the form 
$$I=I^1\times I^2\subseteq X_\inp\times X_\out$$
These contracts capture cases like `inputs are always in range $I^1$ and outputs are always in range $I^2$, independently from one another'.\Hair\footnote{These independent contracts in reality are even more special than that: not only are input restrictions separate from output restrictions, but also each individual \emph{wire} has an associated subset of allowed values on it.} 
Of course this is only a special case of arbitrary relations $R\subseteq X_\inp\times X_\out$, since not all subsets of cartesian products are cartesian products of subsets, as a simple argument in the finite case shows: $|\mathcal{P}(X_\inp\times X_\out)|=2^{n\cdot m}$ whereas $|\mathcal{P}(X_\inp)|\cdot|\mathcal{P}(X_\out)|=2^{n+m}.$ For example, the contract $\{(x,y)\;|\;x<y\}\subseteq \mathbb{R}\times\mathbb{R}$ is not independent.

One could expect that these contracts form themselves an algebra, namely any wiring composite of independent contracts will also be an independent, rather than a general contract itself. However this is not the case in general: although the parallel placement of boxes with $I_X=I_X^1\times I_X^2\subseteq X_\inp\times X_\out$ and $I_Y=I_Y^1\times I_Y^2\subseteq Y_\inp\times Y_\out$ produces the independent contract $(I^1_X\times I^1_Y)\times(I^2_X\times I^2_Y)$ on $X\otimes Y$, closure under feedback fails. Explicitly, for an independent contract $I_X^1\times I_X^2\subseteq X_\inp\times X_\out$ on $X$, and a wiring diagram $(f_\inp,f_\out)\colon X\to Y$, the formula \cref{eq:compositecontract} produces the slightly simpler composite contract
\begin{equation}\label{eq:independentcomposite}
R_Y=\{(y_1,y_2)\in Y_\inp\times Y_\out\;|\;\exists x_2\in I^2_X \textrm{ such that }f_\inp(y_1,x_2)\in I_X^1\textrm{ and }f_\out(x_2)=y_2\}
\end{equation}
which shows that $y_1$ and $y_2$ are not independent in general, hence $R_Y$ is not of the form $I^1_Y\times I^2_Y$.

Notice that in certain examples, $R_Y$ can indeed be written as a product itself, %
for example, when $(f_\inp,\pi_2)$ is of the form $k\times s$ for two functions $k$, $s$. Even more interestingly, due to the special form of morphisms in the wiring diagram category (where they are only made up from projections, diagonals and duplications) in our examples below we will be able to write $R_Y$ as an independent contract itself.\Hair\footnote{It can be shown that independent contracts indeed form an algebra on $\ca{W}$ due to the special morphisms that generate it; the proof is beyond the scope of this paper.}

\subsubsection{Relation to Assume-Guarantee Contracts}

Systems theory and design has long recognized the need for a formal requirement engineering through mathematical models and formal analysis techniques~\cite{Contractsforsystemdesign}. As part of contract-based design, there have been multiple efforts to formalize and analyze \emph{assume-guarantee} (\textsc{ag}) contracts~\cite{romeo:2018} and incorporate them in the design as a fundamental concept. We here discuss such examples and how they fit to the previously described static contract model.

Given a box $\SmallBoxTwo{\mathbb{R}}{\mathbb{R}}{\mathbb{R}}{}$, an example of an assume-guarantee contract (adapted from Benveniste et al.~\cite[\S~IV]{Contractsforsystemdesign}) is \begin{equation}\label{eq:contract1}
R_1:\begin{cases}
\textrm{variables: } & \textrm{ inputs }x,y;\textrm{ outputs }z \\
\textrm{types: } & x,y,z\in\mathbb{R} \\
\textrm{assumptions: } & y\neq0 \\
\textrm{guarantees: } & z=\dfrac{x}{y}
\end{cases}
\end{equation}
This explicitly makes the assumption that the environment (namely the inputs coming either from the external world or from other component systems) will never provide the input $y=0$, essentially leaving the behavior for that input undefined. %
In our formalism, we can express this contract as 
\begin{displaymath}
R_1=\{(x,y,z)\;|\;y\neq0\wedge z=\frac{x}{y}\}\subseteq\mathbb{R}\times\mathbb{R}\times\mathbb{R}
\end{displaymath}
indicating the fact that the input $y=0$ will never occur on the input wire of the box; and if it did, the contract is violated.
A different choice we could make, assuming the initial \textsc{ag} contract is really expressing an "if-then" requirement, is 
\begin{displaymath}
R'_1=\{(x,y,z)\;|\;y\neq0\Rightarrow z=\frac{x}{y}\}\subseteq\mathbb{R}\times\mathbb{R}\times\mathbb{R}
\end{displaymath}
which is a different subset of allowable values on the wires. For example, $(3,0,25)\in R_1'$ whereas $(3,0,25)\notin R_1$.

We now consider a standard \textsc{ag} contract operator called \emph{contract composition and system integration}, and we realize it from the perspective of the wiring diagram algebra machinery -- consequently a more general setting. %
Explicitly, the \textsc{ag} contract composition operator as described for example by Benveniste et al.~\cite[\S~IV.B]{Contractsforsystemdesign} or Le et al.~\cite{le:2016}, takes two \textsc{ag} contracts $R_1=(A_1,G_1)$ and $R_2=(A_2,G_2)$ and produces a new \textsc{ag} contract $R_1\otimes R_2$ (notice that this is a completely different use of our earlier monoidal product symbol $\otimes$) with
\begin{gather}
    G_{R_1\otimes R_2}=G_1\wedge G_2 \label{eq:Benvistecomp}\\
    A_{R_1\otimes R_2}=\textrm{max}\{A\;|\;A\wedge G_2\Rightarrow A_1, A\wedge G_1\Rightarrow A_2\} \nonumber
\end{gather}
only when $R_1$ and $R_2$ are \emph{compatible}, namely $A_{R_1\otimes R_2}\neq\emptyset$. Since $A_{R_1\otimes R_2}$ is the weakest assumption such that the two referred implications hold, if non-empty it ensures that there exists some environment in which the two contracts properly interact: when put in the context of a process that satisfies the first contract, the assumption of the second contract will be met and vice-versa. 
At first sight, this definition looks `symmetric', since it considers a certain compatibility of output guarantee/input assumption in both directions, but in reality this is not quite the case. 

One issue with the above \textsc{ag} contract composition is that the \emph{names} of the variables and not only the types of the wires need to match, in order to connect along them~\cite{Contractsforsystemdesign,nuzzo:2015}. For example, the contract $R_1$ as in \cref{eq:contract1} can be composed with the contract on $\SmallBox{\mathbb{R}}{\mathbb{R}}{}$
\begin{displaymath}
R_2:\begin{cases}
\textrm{variables: } & \textrm{ inputs }u;\textrm{ outputs }x \\
\textrm{types: } & u,x\in\mathbb{R} \\
\textrm{assumptions: } & \top \\
\textrm{guarantees: } & x>u
\end{cases}
\end{displaymath}
not along any wire, as could be deduced by noticing that all wire types are $\mathbb{R}$, but specifically along the wire with variable name $x$. Pictorially, we can realize them as inhabiting boxes wired as
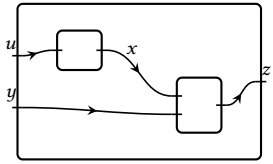
\begin{displaymath}
\begin{tikzpicture}[oriented WD, bbx = .5cm, bby =.8ex, bb min width=.5cm, bb port length=2pt, bb port sep=2]
\node[bb={1}{1}] (X) {$\phantom{X_2}$};
\node[bb={2}{1},below right=.1cm and 1cm of X] (Y) {$\phantom{X_1}$};
\node[bb={2}{1},fit=(X)(Y),inner sep=10pt,inner xsep=15pt] (A) {};
\draw[ar] (X_out1) to (Y_in1);
\draw[ar] (A_in1') to (X_in1);
\draw[ar] (A_in2') to (Y_in2);
\draw[ar] (Y_out1) to (A_out1');
\draw[label]
 node [right=.1in of X_out1] {$\scriptstyle x$}
node [above left= of A_in1] {$\scriptstyle u$}
node [above right= of A_out1] {$\scriptstyle z$}
node [above left= of A_in2] {$\scriptstyle y$}
;
\end{tikzpicture}
\end{displaymath}
and using the formulas \cref{eq:Benvistecomp} we obtain
\begin{gather*}
A_{R_1\otimes R_2}=\textrm{max}\{A\;|\;(A\wedge(x>u)\Rightarrow y\neq0)\wedge(A\wedge (z=x/y)\Rightarrow\top)\}=(y\neq0) \\ G_{R_1\otimes R_2}=(x>u)\wedge(z=x/y).
\end{gather*}
On the other hand, composing $R_1$ and $R_2$ using the static contract algebra (\cref{sec:staticcontracts}) for the above wiring diagram $(f_\inp(x,z,u,y)=(u,x,y),f_\out(x,z)=z)$, we obtain the composite contract
\begin{displaymath}
R=\{(u,y,z)\in\mathbb{R}^3\;|\;\exists x\in\mathbb{R}\textrm{ such that }y\neq0\wedge x>u\wedge z=x/y\},
\end{displaymath}
which could be written in \textsc{ag} form as $A=\{(u,y)\;|\; y\neq0\}$ and $G=\{z\;|\;\exists x>u\textrm{ such that }z=x/y\}$. 
Notice that the contract algebra machinery does not present this variable-match problem, since it does not prevent us from composing along the second input wire of $X_1$ or even do first $X_1$ and then $X_2$ in the opposite order, since all types of wires are real numbers. In all these cases, it would be possible to compute appropriate composite contracts in this uniform way.

The second issue, %
which can also be noticed from the above calculation, is that the assumptions and guarantees of the composite contract include information that mix the variables of the resulting input and output wires. For example, using the \textsc{ag} formalism, the variables of $R_1\otimes R_2$  are set to be $\{u,y\}$ for inputs and $\{x,z\}$ for outputs, %
therefore this operation behaves as if the intermediate wires of a system composition can be extracted as extra output wires to the outside world:
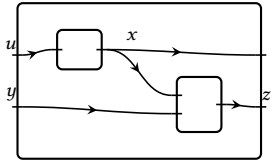
\begin{displaymath}
\begin{tikzpicture}[oriented WD, bbx = .5cm, bby =.8ex, bb min width=.5cm, bb port length=2pt, bb port sep=2]
\node[bb={1}{1}] (X) {$\phantom{X_2}$};
\node[bb={2}{1},below right=.1cm and 1cm of X] (Y) {$\phantom{X_1}$};
\node[bb={2}{2},fit=(X)(Y),inner sep=10pt,inner xsep=15pt] (A) {};
\draw[ar] (X_out1) to (Y_in1);
\draw[ar] (A_in1') to (X_in1);
\draw[ar] (A_in2') to (Y_in2);
\draw[ar] (Y_out1) to (A_out2');
\draw[ar] (X_out1) to (A_out1');
\draw[label]
 node [above right= and .1in of X_out1] {$\scriptstyle x$}
node [above left= of A_in1] {$\scriptstyle u$}
node [above right= of A_out2] {$\scriptstyle z$}
node [above left= of A_in2] {$\scriptstyle y$}
;
\end{tikzpicture}
\end{displaymath}
This `choice' does not agree with the wiring diagram formalism, and moreover is somewhat ad-hoc given that it could potentially add arbitrary many wires to the composite system, essentially according to the result of the contract composition. Adding extra wires is of course possible for the algebra formalism, but corresponds to a choice of architecture on how we decide to wire the subcomponents together, rather than a necessity that arises from dealing with contracts.

Finally, the \textsc{ag} formalism asks that compositions $(R_1\otimes R_2)\otimes R_3$ and $R_1\otimes (R_2\otimes R_3)$ give equivalent contracts, and that so do $R_1\otimes R_2$ and $R_2\otimes R_1$. In the contract algebra formalism, the first statement follows for any $\ca{W}$-algebra: consider a possible wiring of three boxes, each inhabited with a contract (or a behavior)
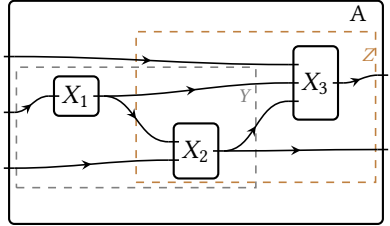
\begin{displaymath}
\begin{tikzpicture}[oriented WD, bbx = .5cm, bby =.8ex, bb min width=.5cm, bb port length=2pt, bb port sep=2]
\node[bb={1}{1}] (X) {$X_1$};
\node[bb={2}{1},below right=.1cm and 1cm of X] (Y) {$X_2$};
\node[bb={3}{1},above right=.05cm and 1cm of Y] (Z) {$X_3$};
\node[bb={3}{2},fit=(X)(Y)(Z),inner sep=17pt] (A) {};
\node[fit=(X)(Y),draw,dashed,gray] {};
\node[fit=(Y)(Z),draw,dashed,yshift=2,brown] {};
\draw[ar] (X_out1) to (Y_in1);
\draw[ar] (A_in1') to (Z_in1);
\draw[ar] (A_in2') to (X_in1);
\draw[ar] (A_in3') to (Y_in2);
\draw[ar] (Y_out1) to (Z_in3);
\draw[ar] (Z_out1) to (A_out1');
\draw[ar] (X_out1) to (Z_in2);
\draw[ar] (Y_out1) to (A_out2');
\draw[label] node at (4.5,0) {$\color{gray}Y$}
node at (7.8,4.5) {$\color{brown}Z$}
node at (7.5,9) {\small A}
;
\end{tikzpicture}
\end{displaymath}

First composing the contracts $R_1$ and $R_2$ and then the result with $R_3$ comes from the application of the functor $\mathcal{C}\colon\ca{W}\to\Cat$ on a wiring diagram morphism
\begin{displaymath}
(X_1\otimes X_2)\otimes X_3\to Y\otimes X_3\to A
\end{displaymath}
whereas the other way around comes from the application of the functor $C$ on the morphism
\begin{displaymath}
X_1\otimes (X_2\otimes X_3)\to X_1\otimes Z\to A
\end{displaymath}
which both express the same morphism $X_1\otimes X_2\otimes X_3\to A$ in $\ca{W}$ as an implementation of $A$ (\cref{sec:theory:arch}).

Regarding the second statement about $R_1\otimes R_2$ and $R_2\otimes R_1$, in the \textsc{ag} formalism this can indeed be proved due to the symmetric formulation of composition \cref{eq:Benvistecomp} as observed earlier. However, this refers more to the earlier variable-sharing clause (which would not allow the composition along arbitrary wires therefore in arbitrary order) and less to composition intuition: changing the order of two boxes and expecting the same behavior or requirements is something highly non expected, from a categorical but also a design  point of view due to the input-output directionality. As a result, commutativity in this \textsc{ag} setting is slightly misleading, since it is just a technical term relevant to the constructed formula (it does not really have an effect on the operation) rather to an actually commuting composition which is not expected to hold -- and does not, in the algebra formalism.

\section{Compositional Cyber-Physical Systems Modeling and Analysis}

In this section, we use the preceding algebraic formalism
to illustrate a compositional \textsc{cps} theory.
We model an unmanned aerial vehicle (UAV),
analyze it with respect to its control behavior,
decompose it to a system architecture
and constrain it using contracts.
This process manifests the power, flexibility and further potential of the wiring diagram compositional framework in the concrete context of \textsc{cps} analysis and design.

\subsection{System Behavior}\label{sec:systembeh}

We algebraically recover a standard controls model \emph{compositionally} in the behavior algebra (Section~\ref{sec:systemsasalgebras}) of the form
\[s_{k+1} = \mathscr{A}s_k + \mathscr{B} c_k,\] 
where $s_k \in \mathbb{R}^n$ is the discrete time state, $s_{k+1}$ (also denoted  $\stackrel{\bullet}{s}$ or $u(s,c)$ using the earlier update function notation) is the subsequent time-step state and $c_k \in \mathbb{R}^n$ is the control signal/output, and
\[y_{k} = \mathscr{C}s_k + \mathscr{D}c_k\]
is the measurement, which is also in $\mathbb{R}^n$. We assume $\mathscr{D} = 0$.

\begin{figure}[!t]
\centering

\begin{tikzpicture}[oriented WD, bbx = .5cm, bby =.8ex, bb min width=.5cm, bb port length=2pt, bb port sep=1]
\node[bb={2}{1}] (L) {\begin{tabular}{c}
     Sensor  \\
     ${L}$
\end{tabular}};
\node[bb={2}{1}, below right= 0.01cm and 1cm of L] (C) {
\begin{tabular}{c}
     Controller\\ ${C}$
\end{tabular}};
\node[bb={1}{1}, above right= 0.01cm and 1cm of C] (D) { \begin{tabular}{c}
     Dynamics\\ ${D}$
\end{tabular}};
\node[bb={2}{1}, fit={($(L.west)-(1,0)$) ($(D.east)+(1,0)$) ($(L.north)+(0,5)$) ($(C.south)-(0,3)$)},gray] (A) {};
 \draw[ar] (A_in1') to (L_in2);
  \draw[ar] (A_in2') to (C_in2);
 \draw[ar] (L_out1) to (C_in1);
 \draw[ar] (C_out1) to (D_in1);
 \draw[ar] (D_out1) to (A_out1');
  \draw[ar] let \p1=(D.north east), \p2=(L.north west), \n1={\y1+2*\bby},  \n2=\bbportlen
  in (D_out1) to [in=0] (\x1-\n2,\n1) -- (\x2-\n2,\n1) to[out=180] (L_in1);
 \node (text) [below=8 of C,rectangle,align=left]
{$ f_\inp\colon\mathbb{R}^3\times\mathbb{R}^2\to\mathbb{R}^5,\; (s',c,s,e,d)\mapsto(s,e,s',d,c)$ \\
$ f_\out\colon\mathbb{R}^3\to \mathbb{R},\;(s',c,s)\mapsto s$};
\node[bb={2}{1},right=6 of D] (L') {$L$}; 
\node[bb={2}{1},below=1 of L'] (C') {$C$};
\node[bb={1}{1},below=1 of C'] (D') {$D$};
\node[dashed,fit=(L')(D'),draw,inner sep=2] {};
\node[bb={2}{1},fit=(L')(D'),inner ysep=23, inner xsep=26] (A') {};
\draw[ar] (A'_in1') to (L'_in2);
  \draw[ar] (A'_in2') to (C'_in2);
 \draw[ar] let \p1=(L'.north east), \p2=(L'.north west), \n1={\y1+2*\bby},\n2=\bbportlen in (L'_out1) to [in=0] (\x1+\n2,\n1) -- (\x2-\n2,\n1) to [out=180] (C'_in1);
\draw[ar] (D'_out1) to (A'_out1');
\draw[ar] let \p1=(L'.north east), \p2=(L'.north west), \n1={\y1+4*\bby},  \n2=\bbportlen in (D'_out1) to [in=0] (\x1+\n2,\n1) -- (\x2-\n2,\n1) to[out=180] (L'_in1);
 \draw[ar] let \p1=(D'.south east), \p2=(D'.south west), \n1={\y1-2*\bby},\n2=\bbportlen in (C'_out1) to [in=0] (\x1+\n2,\n1) -- (\x2-\n2,\n1) to [out=180] (D'_in1);
\draw[label]
node[left=.1 of A_in1] {$\mathbb{R}$}
node[left=.1 of A_in2]{$\mathbb{R}$}
node[right=.1 of A_out1]{$\mathbb{R}$}
node[above right= 7 and 3 of L_out1]{$\mathbb{R}$}
node[above right= 3 and 1 of C_out1]{$\mathbb{R}$}
node[below right= 1 and 1 of L_out1]{$\mathbb{R}$}
node[above right= of A_in1']{$e$}
node[above right= of A_in2']{$d$}
node[above right= of L_out1]{$s'$}
node[below right= of C_out1]{$c$}
node[below right= of D_out1]{$s$}
node[below right= 1 and 6 of C]{UAV}
node[above left= of A'_in1]{$e$}
node[above left= of A'_in2]{$d$}
node[right=.2 of L'_out1]{$s'$}
node[right=.2 of C'_out1]{$c$}
node[above right= of A'_out1']{$s$};
 \end{tikzpicture}

\caption{The physical decomposition of the \textsc{uav}, where $d$ denotes the desired state, $s'$ the predicted state, $c$ the control action, $s$ the current state, and $e$ the environmental inputs.}
\label{fig:behavior_abstract}
\end{figure}
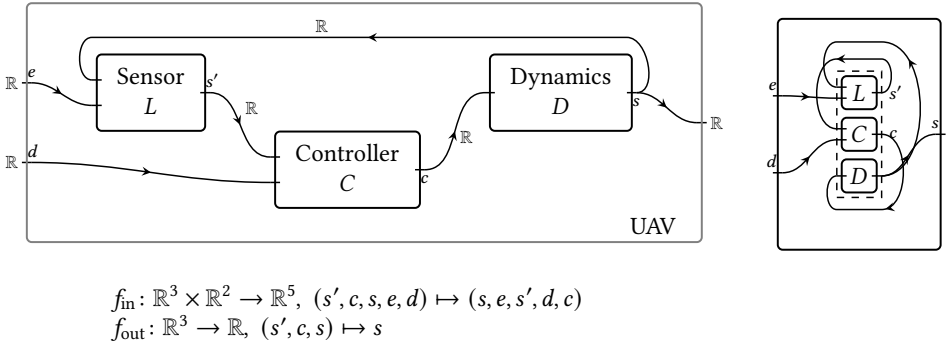

We are going to illustrate the algebra machinery using longitudinal equations of motion for a fixed-winged aircraft represented in the following state-space model~\cite{systemModel}
\begin{equation}\label{eq:boeing}
 \begin{pmatrix}
  \stackrel{\bullet}{a} \\
  \stackrel{\bullet}{q} \\
  \stackrel{\bullet}{\theta}
 \end{pmatrix}=
 \begin{pmatrix}
-0.313 & 56.7 & 0 \\
-0.0139 & -0.426 & 0 \\
0 & 56.7 & 0
 \end{pmatrix}
 \begin{pmatrix}
a \\
q \\
\theta
 \end{pmatrix}+
 \begin{pmatrix}
0.232 \\
0.0203 \\
0
 \end{pmatrix}\begin{pmatrix}\delta\end{pmatrix}
\end{equation}
\begin{displaymath}
 y=\begin{pmatrix}0 & 0 & 1\end{pmatrix}\begin{pmatrix}a \\ q \\ \theta \end{pmatrix},
\end{displaymath}
where $a$ is the angle of attack, $q$ is the pitch rate, $\theta$ is the pitch angle and $\delta$ is the elevator deflection angle. This behavior is the \emph{composite} one, built up from the subcomponents behavior and their wiring (Fig.~\ref{fig:behavior_abstract}).

Working with the linear time-invariant system algebra $\mathcal{L}\colon \ca{W}_\Lin\to\Cat$ (\cref{sec:LTIS}), suppose \((S_L,\mathscr{A}_L,\mathscr{B}_L,\mathscr{C}_L)\), \((S_C,\mathscr{A}_C,\mathscr{B}_C,\mathscr{C}_C)\) and \((S_D,\mathscr{A}_D,\mathscr{B}_D,\mathscr{C}_D)\) are three linear systems inhabiting the respective boxes of Fig.~\ref{fig:behavior_abstract}, with
\begin{align*}
u_L(s_L,s,e)=&\mathscr{A}_L\cdot s_L+ \mathscr{B}_L\cdot(s,e) & 
r_L(s_L)=&\mathscr{C}_L\cdot s_L \\
u_C(s_{C},d,s')=&\mathscr{A}_C\cdot s_C + \mathscr{B}_C\cdot(d,s') &
r_C(s_C)=&\mathscr{C}_C\cdot s_C \\
u_D(s_D,c)=&\mathscr{A}_D\cdot s_D+ \mathscr{B}_D\cdot(c) &
r_D(s_D)=&\mathscr{C}_D\cdot s_D.
\end{align*}
Using the algebra machinery for the specific wiring diagram (Fig.~\ref{fig:behavior_abstract}) given by matrix transformations
\begin{displaymath}
\begin{cases}
f_\inp=\begin{pmatrix}
0 & 0 & 1 & 0 & 0 \\
0 & 0 & 0 & 1 & 0 \\
1 & 0 & 0 & 0 & 0 \\
0 & 0 & 0 & 0 & 1 \\
0 & 1 & 0 & 0 & 0
\end{pmatrix}=\begin{pmatrix}
{}_5(\mathscr{A}^f)_3 & {}_5(\mathscr{B}^f)_2
\end{pmatrix} \\
f_\out=\begin{pmatrix}
0 & 0 & 1
\end{pmatrix}=\mathscr{C}^f
\end{cases}
\end{displaymath}
we can compute the composite linear dynamical system that inhabits the box UAV from the formulas \cref{eq:LTIScomposite}. Its state space is $S_L\times S_{C}\times S_{D}$, and its update and readout linear  functions are
\begin{align*}
u_{\text{UAV}}\colon  S_{L}\times S_{C}\times S_{D}\times\RR^2&\to S_{L}\times S_{C}\times S_{D}, \\
 (s_{L},s_{C},s_{D},d,e)&\mapsto(\mathscr{A}_{L}s_{L}+\mathscr{B}_{L}\begin{pmatrix} \mathscr{C}_{D}s_{D} \\ e \end{pmatrix}, \mathscr{A}_{C}s_{C}+\mathscr{B}_{C}\begin{pmatrix} \mathscr{C}_{L}s_{L} \\ d\end{pmatrix},\mathscr{A}_{D}s_{D}+\mathscr{B}_D\mathscr{C}_{C}s_{C}) \\
r_{\text{UAV}}\colon  S_{L}\times S_{C}\times S_{D}&\to \RR \\
 (s_{L},s_{C},s_{D})&\mapsto\mathscr{C}_{D}s_{D}.
\end{align*}
We assume, for simplicity,\Hair\footnote{See end of this section for a concrete example where $L$ and $C$ are populated by linear functions, thus their state space matches their input linear space  (Section \ref{sec:functions}).} that the state spaces of the sensor and controller are in $\mathbb{R}^2$. Knowing that only the dynamics ${D}$ actually relate to the triplet $(a,q,\theta)$, we deduce that $S_{D}$ is in $\mathbb{R}^3$ which results in a composite state space $S_{\text{UAV}}$ in $\mathbb{R}^2\times\mathbb{R}^2\times\mathbb{R}^3\cong\mathbb{R}^7$.
Moreover, from the shape of the boxes according to \cref{eq:dimensions} we deduce that the matrices $\mathscr{A}_{L}$, $\mathscr{A}_{C}$, $\mathscr{B}_{L}$ and $\mathscr{B}_{C}$ are two-by-two, $\mathscr{C}_{L}$ and $\mathscr{C}_{C}$ are one-by-two, whereas $\mathscr{A}_{D}$ is three-by-three, $\mathscr{B}_{D}$ is three-by-one and $\mathscr{C}_{D}$ is one-by-three. 

Unravelling the above update and readout functions of the composite linear time-invariant system denoted by UAV, the only output of the composite system behavior is that of the dynamics ${D}$, since by tuple \cref{eq:LTIScomposite}
\begin{displaymath}
\mathscr{C}_{\text{UAV}}=\mathscr{C}^f\cdot\mathscr{C}_{L\otimes C\otimes D}=\begin{pmatrix}0 & 0 & _1(\mathscr{C}_{D})_3
\end{pmatrix}.
\end{displaymath}
Hence for obtaining equation \cref{eq:boeing}, in the specific example we deduce that $\mathscr{C}_{D}=\begin{pmatrix}0 & 0 & 1
\end{pmatrix}$ meaning only $\theta$ is output to the outside world as desired. 

For an element of the state space $\RR^7$ of the form $(\vec{s}_{L},\vec{s}_{C},\overbrace{a,q,\theta}^{\vec{s}_{D}})$, isolating the first two variables we obtain
\begin{displaymath}
\stackrel{\bullet}{\vec{s}_{L}}=\mathscr{A}_{L}\vec{s}_{L}+{}_2(\mathscr{B}_{L})_2\begin{pmatrix} \overbrace{\mathscr{C}_{D}\vec{s}_{D}}^\theta \\ e \end{pmatrix}\text{\quad and \quad }
\stackrel{\bullet}{\vec{s}_{C}}=\mathscr{A}_{C}\vec{s}_{C}+{}_2(\mathscr{B}_{C})_2\begin{pmatrix}\mathscr{C}_{L}\vec{s}_{L} \\ d \end{pmatrix},
\end{displaymath}
which could be viewed as some extra information of the composite system relating to the behaviors of the sensor and controller, not appearing in equation \cref{eq:boeing} but part of the total system's behavior. %

Now isolating the last three variables we obtain a description
\begin{displaymath}
\begin{pmatrix}\stackrel{\bullet}{\alpha} \\
\stackrel{\bullet}{q} \\
\stackrel{\bullet}{\theta}
\end{pmatrix}=
{}_3(\mathscr{A}_{D})_3\begin{pmatrix}\alpha \\
q \\
\theta
\end{pmatrix}+{}_3(\mathscr{B}_{D})_1\mathscr{C}_{C}\vec{s}_{C}.
\end{displaymath}
Comparing with the desired equation \cref{eq:boeing}, the elevator deflection angle $\delta$ is the output of the controller $\mathscr{C}_{C}s_{C}$ which  matches the physical reality, and the $\mathscr{A}_{D}$, $\mathscr{B}_{D}$ are completely determined by the composite description, namely
\begin{displaymath}
\mathscr{A}_{D}= \begin{pmatrix}
-0.313 & 56.7 & 0 \\
-0.0139 & -0.426 & 0 \\
0 & 56.7 & 0
 \end{pmatrix}
\qquad
\mathscr{B}_{D}= \begin{pmatrix}
0.232 \\
0.0203 \\
0
 \end{pmatrix}.
\end{displaymath}
The remaining data $\mathscr{A}_{{L},{C}},\mathscr{B}_{{L},{C}},\mathscr{C}_{{L},{C}}$ depend on engineering and physical parameters. 

We were thus able to partly reverse-engineer a given composite system behavior \cref{eq:boeing}, where for the given system architecture (Fig.~\ref{fig:behavior_abstract}) we completely identified the behavior of the linear time-invariant system $D$ by determining $S_D,\mathscr{A}_D,\mathscr{B}_D,\mathscr{C}_D$. We also obtained certain information about the other two subcomponents $C$ and $L$: for example, two possible behaviors could be the linear functions (for example, signal concatenations) $s'=s+e$ for the sensor $L$ and the linear function $c=s'+d$ for the controller $C$. %
Expressing those as linear time-invariant systems (\cref{sec:functions}), we obtain the following description
\begin{gather*}
(S_L,\mathscr{A}_L,\mathscr{B}_L,\mathscr{C}_L)=\left(\mathbb{R}^2,\begin{pmatrix}
0 & 0 \\
0 & 0
\end{pmatrix},
\begin{pmatrix}
1 & 0 \\
0 & 1
\end{pmatrix},
\begin{pmatrix}
1 & 1
\end{pmatrix}\right),\qquad u_L(\vec{s}_L,s,e)=(s\;\;e),\; r_L(\vec{s}_L)=s_L^1+s_L^2 \\
(S_C,\mathscr{A}_C,\mathscr{B}_C,\mathscr{C}_C)=\left(\mathbb{R}^2,\begin{pmatrix}
0 & 0 \\
0 & 0
\end{pmatrix},
\begin{pmatrix}
1 & 0 \\
0 & 1
\end{pmatrix},
\begin{pmatrix}
1 & 1
\end{pmatrix}\right),\qquad u_C(\vec{s}_C,s',d)=(s'\;\;d),\; r_C(\vec{s}_C)=s_C^1+s_C^2
\end{gather*}
Then the composite system's update function is explicitly computed, using \cref{eq:LTIScomposite}, as
\begin{displaymath}
\begin{cases}
\stackrel{\bullet}{\vec{s}}_L=\begin{pmatrix}
\theta \\
e
\end{pmatrix} \\
\stackrel{\bullet}{\vec{s}}_C=\begin{pmatrix}
s^1_L+s^2_L \\
d
\end{pmatrix} \\
\stackrel{\bullet}{a}=-0.313a+56.7q+0.232s_C^1+0.232s_C^2 \\
\stackrel{\bullet}{q}=-0.0139a-0.426q+0.0203s_C^1+0.0203s_C^2 \\
\stackrel{\bullet}{\theta}=56.7q
\end{cases}
\end{displaymath}
where $s_C^1$ and $s_C^2$ are essentially the previous desired state $s'$ and input $d$, producing the deflection angle $\delta$ that appears in \cref{eq:boeing}.
The first two equations give the functions of $L$ and $C$ (whose states are placeholders for their inputs at each instance), whereas the last three give the dynamics $D$ as before.
Informally, this shows the interplay between what the system is sensing, what its desired operating state is, and how it must react. If there were more information about the elevator deflection angle $\delta$, that would restrict the possible behaviors for $C$ appropriately. %

From a more categorical perspective, the above process is summarized as follows: given an algebra $\mathcal{L}$ and a wiring diagram $f\colon{L}\otimes{C}\otimes{D}\to \text{UAV}$ in $\ca{W}_\Lin$ (Fig.~\ref{fig:behavior_abstract}), as well as an object of the target category $\mathcal{L}({\text{UAV}})$, namely a specific linear system as in equation \cref{eq:boeing} inhabiting the outside box $\text{UAV}$, the goal is to find an object in the pre-image of the given system under the composite functor
\begin{displaymath}%
 \mathcal{L}({L})\times \mathcal{L}({C})\times \mathcal{L}({D})\xrightarrow{\phi_{{L},{C},{D}}}
 \mathcal{L}({L}\otimes{C}\otimes{D})\xrightarrow{\mathcal{L}(f)}\mathcal{L}(\text{UAV})\text{.}
\end{displaymath}
Such a problem certainly does not have a unique solution, namely a unique description of the three systems that form the composite, but for example in this specific case due to the form the wiring diagram, the component system 
\begin{displaymath}%
(S_{D},\mathscr{A}_{D},\mathscr{B}_{D},\mathscr{C}_{D})    
\end{displaymath}
was completely determined by the composite behavior. Further work would aim to shed light on possible shapes of wiring diagrams that have better identifiable solutions under algebras of interest.

\subsection{System Architecture}\label{sec:systemarch}

One of the important advantages of expressing system decompositions as a morphism in the category $\ca{W}$ is that we can perform further zoomed-in decompositions as desired in a \emph{hierarchical} way (\cref{sec:theory:arch}), and these are all realized as composite morphisms in the wiring diagram category.

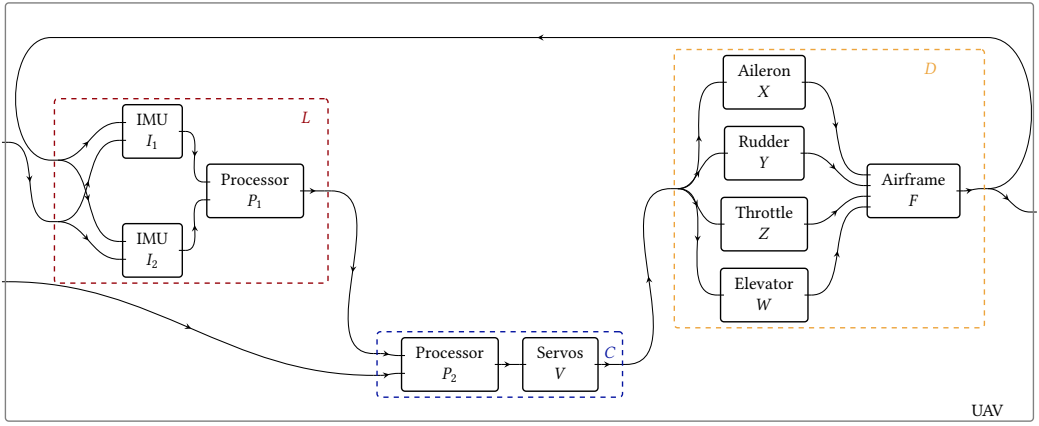
\begin{figure}[!t]
\centering

\begin{tikzpicture}[oriented WD, bbx = .5cm, bby =.8ex, bb min width=.5cm, bb port length=2pt, bb port sep=1]
\node[bb={2}{1}] (P) {{\begin{tabular}{c}
     Processor\\ $P_1$
\end{tabular}}};
\node[bb={2}{1},above left= of P] (I) {\begin{tabular}{c}
     IMU\\ $I_1$
\end{tabular}};
\node[bb={2}{1},below left=of P] (J) {\begin{tabular}{c}
     IMU\\ $I_2$
\end{tabular}};
\node[bb={2}{1},inner sep=.1in,fit={($(J)-(3,0)$)(J)(I)(P)},effectivenessCurve,dashed] (L) {};
\node[bb={2}{1}, below right= 1.1cm and 1.5cm of L] (Q) {{\begin{tabular}{c}
     Processor\\ $P_2$
\end{tabular}}};
\node[bb={1}{1},right= of Q] (V) {{{\begin{tabular}{c}
     Servos\\ $V$
\end{tabular}}}};
\node[bb={2}{1}, inner sep=.1in,fit={(Q)(V)},CostCurve,dashed] (C) {};
\node[bb={1}{1}, above right=0.2cm and 2cm of C] (U) {{\begin{tabular}{c}
     Elevator\\ $W$
\end{tabular}}};
\node[bb={1}{1}, above=1em of U] (Z) {{\begin{tabular}{c}
     Throttle\\ $Z$
\end{tabular}}};
\node[bb={1}{1}, above=1em of Z] (Y) {{\begin{tabular}{c}
     Rudder\\ $Y$
\end{tabular}}};
\node[bb={1}{1}, above=1em of Y] (X) {{\begin{tabular}{c}
     Aileron\\ $X$
\end{tabular}}};
\node[bb={4}{1}, right=11.5cm of P] (F) {{\begin{tabular}{c}
     Airframe\\ $F$
\end{tabular}}};
\node[bb={1}{1}, fit={($(X.west)-(1,0)$)(X)(U)(F)},yel,dashed] (D) {};
\node[bb={2}{1}, fit={($(L.west)-(1,0)$) ($(D.east)+(1,0)$) ($(L.north)+(0,15)$) ($(C.south)-(0,3)$)},gray] (A) {};
\draw[ar] (L_in1') to (I_in1);
\draw[ar] (L_in1') to (J_in1);
\draw[ar] (L_in2') to (J_in2);
\draw[ar] (L_in2') to (I_in2);
\draw[ar] (I_out1) to (P_in1);
\draw[ar] (J_out1) to (P_in2);
\draw[ar] (P_out1) to (L_out1');
 \draw[ar] (A_in2') to (C_in2);
  \draw[ar] (A_in1') to (L_in2);
 \draw[ar] (L_out1) to (C_in1);
 \draw[ar] (C_out1) to (D_in1);
 \draw[ar] (V_out1) to node[above,CostCurve]{$C$} (C_out1');
 \draw[ar] (D_out1) to (A_out1');
\draw[ar] (C_in1') to (Q_in1);
\draw[ar] (C_in2') to (Q_in2);
\draw[ar] (Q_out1) to (V_in1);
\draw[ar] (D_in1') to (X_in1);
\draw[ar] (D_in1') to (Y_in1);
\draw[ar] (D_in1') to (Z_in1);
\draw[ar] (D_in1') to (U_in1);
\draw[ar] (X_out1) to (F_in1);
\draw[ar] (Y_out1) to (F_in2);
\draw[ar] (Z_out1) to (F_in3);
\draw[ar] (U_out1) to (F_in4);
\draw[ar] (F_out1) to (D_out1');
\draw[ar] let \p1=(D.north east), \p2=(L.north west), \n1={\y1+2*\bby}, \n2=\bbportlen in(D_out1) to[in=0] (\x1+\n2,\n1) -- (\x2-\n2,\n1) to[out=180] (L_in1);
\node[below right= .2 and 14 of C]{UAV};
\node[above right= -.5cm and 2.3cm of I,effectivenessCurve]{$L$};
\node[above right= -.5cm and 2.3cm of X,yel]{$D$};
 \end{tikzpicture}

\caption{Any decomposition, including the previous one (Fig.~\ref{fig:behavior_abstract}) resides within the slice category $\ca{W}/\text{UAV}$. In this case, the slice category contains all possible design decisions that adhere to the behavioral model; we pick one such design choice.} \label{fig:full_hier}
\end{figure}

For example, consider a possible UAV architecture (Fig.~\ref{fig:behavior_abstract}).
We may further choose to implement the sensor box ${L}$ using two IMU units ${I}_1,{I}_2$ and a processor ${P}_1$ in a certain interconnection. Expressing this as a morphism with target ${L}$ (an object in the slice category $\ca{W}/L$) namely $g\colon{I}_1\otimes {I}_2\otimes {P}\to {L}$ means that we can compose this with the original one-level implementation $f$ to obtain a two-level zoomed-in decomposition
\begin{displaymath}
({I}_1\otimes{I}_2\otimes{P}_1)\otimes {C}\otimes {D}\xrightarrow{g\otimes1\otimes1}{L}\otimes {C}\otimes {D}\xrightarrow{f}{\text{UAV}}
\end{displaymath}
that only `opens-up' the box ${L}$. We could moreover implement the control as well as the dynamics box, and decompose them in a choice of subcomponents and wires between them. An example where the control box is decomposed into ${P}_2$ followed by ${V}$ in a serial composition, and the dynamics box is decomposed into four parallel boxes, ${X}$, ${Y}$, ${Z}$ and ${W}$ followed by ${F}$ amounts to choosing a specific $h\colon {P}_2\otimes{V}\to {C}$ in $\ca{W}/C$ and a specific $k\colon {X}\otimes{Y}\otimes{Z}\otimes{W}\otimes{F}\to{D}$ in $\ca{W}/D$. Combining all these morphisms we have the composition (Fig.~\ref{fig:full_hier}):
\begin{displaymath}
({I}_1\otimes {I}_2\otimes {P}_1)\otimes ({P}_2\otimes {V})\otimes ({X}\otimes {Y}\otimes {Z}\otimes {W}\otimes {F})\xrightarrow{g\otimes h\otimes k}{L}\otimes {C}\otimes {D}\xrightarrow{f}{\text{UAV}}
\end{displaymath}
that can be considered as a single morphism from the tensor of all second-level sub-components to the box UAV. Pictorially, this would be realized by erasing the intermediate colored dashed boxes.

\subsection{System Requirements}\label{sec:systemcontr}

We will use the algebra of static contracts $\mathcal{C}\colon\ca{W}\to\Cat$  (\cref{sec:staticcontracts}), where all requirements are expressed as subsets of the cartesian product of input and output types. 
Consider the original system decomposition to sensor, controller, and dynamics boxes (Fig.~\ref{fig:behavior_abstract}) and suppose we have certain contracts on these components given by
\begin{gather*}
R_{L}\subseteq\RR^2\times\RR, \quad R_{C}\subseteq\RR^2\times\RR, \quad
R_{D}\subseteq\RR\times\RR.
\end{gather*}
These contracts could be any subsets, from the extreme case of equality which means that \emph{all} combinations of inputs and outputs are allowed, to some specific requirement imposed to the example at hand, or in certain cases some \emph{maximal} contract dictated by a discrete dynamical system (governed by a difference equation) that actually inhabits the box. %

The contract algebra applies to the wiring diagram of Fig. \ref{fig:behavior_abstract} and based on the formula \cref{eq:compositecontract} produces a contract $R_\text{{UAV}}\subseteq\mathbb{R}^2\times\mathbb{R}$
on the composite system, with the following explicit description \begin{align*}
\RR^3\supseteq R_\text{{UAV}} = & \{(a_1,a_2,a_3)\in\RR^3\;|\;\exists (x,y)\in\RR^2\textrm{ such that } \\
& (a_3,a_1,x)\in R_L,\; (x,a_2,y)\in R_C,\; (y,a_3)\in R_D\}.
\end{align*}
Further, we could assume that all contracts are independent as per \cref{sec:independentcontr}, namely they can be written as products of subsets of each wire type independently, like
\begin{gather*}
R_{L}=R_{L}^1\times R_{L}^2\times R_{L}^3, \quad
R_{C}=R_{C}^1\times R_{C}^2\times R_{C}^3, \quad
R_{D}=R_{D}^1\times R_{D}^2
\end{gather*}
where all components are subsets of $\mathbb{R}$ -- i.e. the allowed values on each wire are completely unrelated to one another. Then the composite contract (\ref{eq:independentcomposite}) takes the following, also independent contract form 
\begin{equation}\label{eq:RUAV}
R_{\text{UAV}}=
\begin{cases} 
R_L^2\times R_C^2\times(R_L^1\cap R_D^2)
 & \textrm{if }R_L^3\cap R_C^1\neq\emptyset\textrm{ and }R_C^3\cap R_D^1\neq\emptyset
 \\
 \qquad\qquad\emptyset & \textrm{if }R_L^3\cap R_C^1=\emptyset\textrm{ or }R_C^3\cap R_D^1=\emptyset 
\end{cases}
\end{equation}
The above formula expresses that the allowable tuples that can be observed on the composite system are the $L$- and $C$-external input contracts for the two input wires, along with an intersection of contracts for the output wire, subject to whether there exists a scenario where the contracts of the intermediate wires match: if their intersection is non-empty, there exist appropriate values that work for both contracts and the total system `runs'. Otherwise the composed contracts are incompatible and the composite system fails to adhere to a contract, namely there is no guarantee about its observable input and output values (expressed by the empty set contract) and possibly the whole process fails.

We now proceed to a similar process to what we have seen before (\cref{sec:systembeh}), which in a sense reverse-engineered the behavior of the subcomponents, given a composite behavior of the total system using the system behavior algebra machinery. In this setting, given a specific desired requirement $R_\text{UAV}$ on the composite system, we will identify possible contracts on the components that produce that specific composite; once again we do not expect unique solution to this problem.

Suppose the envisioned composite contract on the behavioral representation of our example UAV (Fig.~\ref{fig:behavior_abstract})
is 
\begin{displaymath}
R_{\text{UAV}}=[0,100]\times[-20,+20]\times[-35,+35]
\end{displaymath}
This contract represents a possible requirement that the desired UAV pitch is no more or less than $20$ degrees and the plane really must not pitch more or less than $35$ degrees for a hypothetical safe flight. As hypothetical environmental conditions, we assume air speed does not exceed $100$ \si[per-mode=symbol]{\kilo\metre\per\hour}.

Comparing the above composite contract against equation \cref{eq:RUAV}, we can first of all deduce that 
\begin{displaymath}
R^2_L=[0,100]\qquad R^2_C=[-20,+20]
\end{displaymath}
namely the external inputs for $L$ and $C$ are necessarily constrained by the ranges of the given composite contract on those wires.
Moreover we have that $R^1_L\cap R^2_D=[-35,35]$ and also that necessarily the intersections $R^3_L\cap R^1_C$ and $R^3_C\cap R^1_D$ are non-empty -- since the composite contract is indeed non-empty. Notice how all these intersections correspond to specific wiring connections or splittings we performed between subcomponents for the initial UAV's implementation.

Given these restrictions, we are free to choose contracts that satisfy them, for example
\begin{displaymath}
R^2_D=[-35,+35],\;R^1_L=R^3_L=R^1_C=R^3_C=R^1_D=\mathbb{R}
\end{displaymath}

The above choices are made to also dispose of `bad scenarios' for the given interconnection of the boxes. For example, choosing the opposite contracts for $R^2_D$ and $R^1_L$ would be mathematically correct since their intersection is still $[-35,35]$, but could lead to a real value of, say, $40$ degrees entering the sensor $L$ which would then violate its contract (that said "all my inputs on the first wire will be less than $35$"). Although in general, processes can be wired together as long as types match, in the contract algebra setting it is implied (by the algebra machinery) that the only values passing through an interconnected wire are those in the intersections of the individual (independent) contracts -- so long as the composite system does not `break'. It is important to realize that the contract algebra \emph{describes} the observable inputs and outputs on a running composite machine, rather than \emph{ensures} that the process runs: this has to be safeguarded by the designer also. This discussion relates to future work regarding `total' or `deterministic' contracts. %

\section{On Unification}\label{sec:unification}

Having manifested the wiring diagram formalism for behavior, architecture and requirements of an UAV, we now summarize and further discuss how this categorical interpretation of \textsc{cps} models leads to unification of these aspects of system design and analysis.

Starting with some cyber-physical process $Y$, we usually model its behavior, mathematically described for example via some equations, and also the requirements it satisfies or should satisfy. We earlier discussed Moore machines and linear time-invariant systems; there can be other algebras of system behavior,\Hair\footnote{For example, \emph{machines} serve as an all-inclusive general system notion that allows us to compose systems of different description~\cite[\S~4]{spivak:2016}.} so here we generically speak of the `behavior algebra' which is any one of them, using the notation $\mathcal{B}$. As we saw, categorically these are certain objects $B_Y\in \mathcal{B}(Y)$ of the category of all the possible behaviors (\cref{sec:systemsasalgebras}), and similarly the requirements are objects $R_Y\in \mathcal{C}(Y)$ of the category of all contracts (\cref{sec:contracts}) that could be associated to such a process, via lax monoidal functors
\begin{displaymath}
\mathcal{B},\mathcal{C}\colon\ca{W}\to\Cat.
\end{displaymath}
To formally discuss and capture the behavior and requirements in terms of subprocesses, the designer first chooses some valid architecture of $Y$ which is categorically expressed by choosing a morphism $f\colon X\to Y$ in the category $\ca{W}$, namely an element of the slice category $\ca{W}/Y$ (\cref{sec:theory:arch}). Then the behavior algebra and requirements algebra,
independently, produce assignments
\begin{equation}\label{eq:emptynat}
\begin{tikzcd}[row sep=.05in,column sep=.7in]
& \mathcal{B}(X)\ar[dd,"\mathcal{B}(f)"] \\
& \phantom{A} \\
X\ar[dd,"f"'] & \mathcal{B}(Y) \\
\phantom{B}\ar[uur,bend left,dotted]\ar[ddr,bend right,dotted] & \\
Y & \mathcal{C}(X)\ar[dd,"\mathcal{C}(f)"] \\
& \phantom{C} \\
& \mathcal{C}(Y)
\end{tikzcd}
\end{equation}
The designer then decides on `pre-image' objects $B_X\in \mathcal{B}(X)$ and $R_X\in\mathcal{C}(X)$ which, under these functors on the right-hand side, produce the original composite behavior and requirement on $Y$. As we saw, there could be multiple choices for $B_X$ and $R_X$ (\cref{sec:systembeh,sec:systemcontr}). Also, the designer can decompose even further to subprocesses, on which the analysis carries on in the same formal way (Section \ref{sec:systemarch}). 
Moreover, they may choose to go back and change the architecture to some alternative implementation $g\colon Z\to Y$, if that is physically sensible and allows to easier obtain the end results. Later on, using algorithms such tests could assist in deciding on the most optimal solutions.

On top of the above story, which summarizes the narrative of the current work, we now sketch some additional connections between these two independent algebras of behavior and requirements, which further clarify their formal relation.

First of all, there is an \emph{algebra map}\Hair\footnote{Formally, this is a monoidal natural transformation between the two lax monoidal functors~\cite[\S~XI]{mac:1998}.} $\alpha\colon \mathcal{B}\Rightarrow \mathcal{C}$ which assigns to each specific physical behavior of a process $B_Y\in \mathcal{B}(Y)$, the \emph{maximally satisfied} contract by it, $\alpha_Y(B_Y)\in\mathcal{C}(Y)$; in \cite[Prop.~5.2.15]{spivak:2016} this is done in an abstract setting. Informally, if a box $\SmallBox{\mathbb{R}}{\mathbb{R}}{X}$ is inhabited by the function $f(x)=6x$, its maximally satisfied contract is in effect $\{(a,6a)\;|\;a\in\mathbb{R}\}\subseteq\mathbb{R}^2$. However, the system also satisfies the contracts 
$\mathbb{R}\times6\mathbb{R}$
or $\mathbb{R}\times3\mathbb{R}$, or even $\mathbb{R}\times\mathbb{R}$ as the maximum such. %
The fact that the assignment $\mathcal{B}(Y)\ni B_Y\mapsto \alpha_Y(B_Y)\in\mathcal{C}(Y)$ is an algebra map signifies in particular that the above mappings \cref{eq:emptynat} are part of a commutative square relating system behavior and requirements for a specific wiring diagram $f\colon X\to Y$
\begin{displaymath}
\begin{tikzcd}
\mathcal{B}(X)\ar[d,"\mathcal{B}(f)"']\ar[r,"\alpha_X"] & \mathcal{C}(X)\ar[d,"\mathcal{C}(f)"] \\
\mathcal{B}(Y)\ar[r,"\alpha_Y"'] & \mathcal{C}(Y)
\end{tikzcd}
\end{displaymath}
Intuitively, this says that for a given system decomposition into subcomponents, first composing the behaviors of the internal boxes using the behavior algebra and then talking about the contract that composite satisfies is the \emph{same} as first computing the maximal contracts the components satisfy individually and then composing using the contract algebra. This provides extra flexibility for passing between different models, not only for this specific algebra map example but also for other maps relating different algebras that may be established.

Another way to combine the behavior $\mathcal{B}$ and requirements $\mathcal{C}$ algebra is to construct a new algebra of \emph{contracted behaviors} that, to each process placeholder $Y$ assigns a pair of a physical behavior along with some contract it satisfies. This allows us to compose using both algebras simultaneously and choosing which information to look at; this abstract algebra is already defined \cite[Prop.~4.5.5]{spivak:2016} for a specific behavior algebra and provides a tool that allows us to essentially relate two algebras via some desired condition inside their product.

The above sketched behavior and requirements formal connections, as well as the whole methodology presented in detail in this paper, shall be further developed to account for the crucial notion of time,\Hair\footnote{The abstract categorical framework where time is added to the wiring diagram model has been formally studied using \emph{sheaves} on real-time intervals~\cite[\S~3]{spivak:2016}.}
particularly a \emph{compositional} model
of real-time computing
for \textsc{cps}, which to this day raises several challenges~\cite{stankovic:1988,edwards:2007,lee:2016,lee:2018}.

\section{Related Work}

Computational and physical modeling in the context
of \textsc{cps} is well-studied~\cite{derler:2012,zhu:2018,platzer:2019}.
However, there is still a need for research in compositional methods for model-based system design~\cite{durn:2020,tripakis:2016}
and particularly for a compositional \textsc{cps} theory
that is able to model and simulate both the computational and the physical aspects of \textsc{cps}~\cite{lee:2006,bliudze:19,cremona:2019}, which can formally relate those necessary views.

Category theorists have worked extensively
in the area of compositional systems.
Among the primary results of that general program
has been the \emph{relation}
of different types of models, for example,
abstracting and unifying automata and dynamical systems~\cite{arbib:1980}. For further discussion and references on alternative categorical approaches on systems theory, see Schultz et al.~\cite[\S~1]{spivak:2016}.
Furthermore, the application of algebraic structures has made seminal contributions in behavioral specification
of programs~\cite{bidoit:1995} 
and modal logic~\cite{hughes:2001}, both 
of which show up later in control~\cite{fainekos:2005}.
Category theory has recently been used
in mobility~\cite{zardini:2020,zardini:2020a}, planning and scheduling~\cite{breiner:2020}, robotics~\cite{zardini:2021},
and security assessment~\cite{bakirtzis:2021},
which indicates it is a developing field within engineering.

Other recent work in systems theory proposes category theory
as the solution to model federation 
but lacks significant theoretical development.
Hasuo~\cite{hasuo:2017}, for example, provides much
of the context and reasoning
for using category theory in system design through coalgebras
but the work represents a skeleton of what should be done.
An older but significantly more fleshed out version of coalgebraic modeling for \textsc{cps} was proposed by Matsikoudis and Lee~\cite{matsikoudis:2012}, but only focuses on modeling the behavioral view of transition systems.
Additionally, category theory lends itself as a possible \emph{quasi}-formal approach to requirements management~\cite{gebreyohannes:2018,kibret:2019}.
Our framework is instead formal, in the strict sense,
and we use the contracts algebra, which has shown to be effective
in \textsc{cps}.

The theory of contracts has had significant development, especially as applied to \textsc{cps}~\cite{sangiovanni:2012,benveniste:2018}, including notions of contract composition~\cite{oh:2019}.
Recently there are also concrete applications
in the form, for example, of a toolkit on top of SysML~\cite{dragomir:2017},
which will make contracts increasingly accessible
to system designers.
Contracts have been implemented as an end-to-end requirements engineering framework, but more importantly have also been merged with linear temporal logic (\textsc{ltl}) specifications that can compile down to contracts~\cite{nuzzo:2018b}; this idea could also be implemented into our compositional \textsc{cps} theory.
Our approach to contracts is more general
than the often used \textsc{ag} formalism.
Specifically, in \textsc{ag} contracts, names \emph{and} types of variables need to match, while the categorical formulation only requires that types match. 
Examples of synthesis from a contract-based design specification~\cite{ghasemi:2020}, show that it is possible to use our generalized version of contracts to adapt control synthesis tools~\cite{mazo:2010,rungger:2014} with our notion of modeling and simulation.
Therefore, we would be able to not only have composition among requirements, system behaviors, and system architectures but we would also be able to produce a possible implementation that is compositionally constrained at any given level; this would represent an improvement over approaches that only consider the compositional verification of architecture models~\cite{cofer:2012}.

Hybrid systems is a well-established version
of computational dynamical systems theory~\cite{alur:2000} (another being timed process algebras~\cite{broy:1993} or more recently model interfaces~\cite{raclet:2011}).
Ames (as well as Tabuada et al. to some extent~\cite{tabuada:2002,tabuada:2008}) did develop a categorical theory
of hybrid systems~\cite{ames:2006},
which could be used to relate
or otherwise use results from the well-established formalism
of hybrid systems in our proposed framework as future work,
but also the opposite; potentially strengthening notions of composition~\cite{alur:2006,bresolin:2020}
in hybrid systems from category theory.
These problems have been so far mostly tackled from a theoretical sense categorically without yet directly relating to systems science~\cite{culbertson:2020}. If we desired the behavior to be modeled within the hybrid system paradigm we are able to directly import those results within our framework. In the future, by incorporating this hybrid systems formalism it is possible for the categorical framework to use compositional structures into standard analysis methods, such as reachability analysis~\cite{bogomolov:2020}, modular \textsc{smt}-based hybrid systems analysis~\cite{bae:2017}, and passivity/stability analysis~\cite{koutsoukos:2012}.

The congruence between logic and category theory also means this framework could potentially be used  to augment logic-based approaches to \textsc{cps} design, such as contract composition in the KeYmaera X tool~\cite{muller:2018} or hybrid Event-B~\cite{banach:2017}. The inverse also holds true. Our framework could become richer or manage richer semantics by incorporating a logical framework in addition to the behavioral and requirements semantics presented in this paper while keeping composition explicit within the design problem or further assist with the separation of computation and physics present in \textsc{cps}~\cite{lunel:2019}. The augmentation of compositional \textsc{cps} theory with logic can also combat the problem of specification mismatching throughout the lifecycle of the system and its associated models~\cite{nguyen:2018}.

Compositional \textsc{cps} theory can assist with model conformance~\cite{roehm:2019} and model federation at large~\cite{golra:2018}. Complementary works
using category theory have shown small but useful examples
of categorical systems modeling 
and how they can facilitate model conformance in \textsc{cps}~\cite{breiner:2019a,breiner:2019,nolan:2019,bakirtzis:2020}.

\section{Conclusion}

The forthcoming SysML V2 standard is attempting
to bridge the gap between requirement, behavioral,
and structural models, showing an increasing need
for unification and scalability of models
in system design~\cite{omg:2019}.
In this paper, we present a categorical framework
to achieve such a unification of models
and simulation tools as an alternative 
to current approaches, such as domain metamodeling
and (semi-)manual model transformations.
Through the categorical framework we also show that there is a functorial relationship between the architectural and behavioral modeling domains, which unifies what was previously a distinct difference between these domains.
We show that there is a multidimensionality to modeling abstraction
and manage it formally through the preceding formalism.

An additional benefit of category theory
in this domain is its closeness to execution
by means of dependent typed languages,
which in the future could allow for a merge
between modeling and code that traverses
throughout the full lifecycle
of the system.
This would lead to further unification
in the form of managing algebras of programming languages in addition to the algebras we have implemented in this paper,
thereby allowing for further options in regards
to both models of computation for \textsc{cps}
and flexibility in synthesizing systems from models.
These stages might include requirements generation, control law simulation,
and finally architectural design and deployment.
In the domain of \textsc{cps} we achieve that
by unifying the controls and computation and requirement views
without inventing a new formalism
but rather by zooming in and out of different layers
of abstraction with a formal composition rule.
Furthermore, with this approach we are able
to relate static views of the system
with their dynamics, or otherwise executable, model representations.

The use of wiring diagrams already provides
both an appealing and familiar syntax (that of boxes and arrows)
as well as algebraic semantics; that is, the perspective of \emph{systems as algebras},
which formalizes mathematically the diagrammatic reasoning already used
in engineering.
We posit that as systems become increasingly complex
such semantics will be important to assess
a system's dependable, safe, and secure operation.
These semantics do not need to be visible
to the practitioner but provide a flexible scaffolding
for interchanging between modeling paradigms
and metrics within a modeling language.
Ultimately, the algebraic view
of systems models has the potential of producing more scalable modeling efforts.

\begin{acks}
G. Bakirtzis and C.H. Fleming are partially supported through \textsc{serc} under \textsc{usdod} Contract HQ0034-13-D-0004, \textsc{nasa} under research grant NNX16AK47A, and \textsc{nsf} under grant No. 1739333.

C. Vasilakopoulou is supported by the General Secretariat for Research and Technology (\textsc{gsrt}) and the Hellenic Foundation for Research and Innovation (\textsc{hfri}).
\end{acks}

\bibliographystyle{ACM-Reference-Format}
\bibliography{manuscript}

 \appendix
 \section{Nomenclature}\label{nomenclature}
 
Here we summarize some of the symbols we use and their meaning
in category theory as a quick guide for working engineers
to effectively navigate the preceding formalism.

{\def\arraystretch{1.7}\tabcolsep=10pt
 \begin{longtable}{cl}
$\ca{C}$ & a generic category\\
$1$ & identity morphism\\
$f \text{ or } \xrightarrow[]{f}$ & morphism in a category\\
$g \circ f$ & composition (right to left)\\
\begin{tikzcd}[sep=small,cramped,ampersand replacement=\&]A\arrow[r,"f"]\arrow[d,"h"'] \& B\arrow[d,"g"] \\ C\arrow[r,"k"] \& D\end{tikzcd} & commutative diagram standing for equation $g\circ f=k\circ h$\\
$\cong$ & {isomorphism}\\
$(\ca{V},\otimes,I)$ & a generic monoidal category\\
$F \text{ or } \xrightarrow[]{F}$ & functor\\
$F(A) \text{ or } FA$ & functor application on objects\\
$F(f) \text{ or } Ff$ & functor application on morphisms\\
$\ca{C}/C$ & slice category over object $C$\\
$\Set$ & the category of sets and functions\\
$\ca{Lin}$ & the category of linear spaces and linear maps\\
$\times$ & cartesian product of sets (or linear spaces) \\
$\Delta\colon X{\to}X{\times}X$ & duplication function\\
$\Cat$ & the category of categories and functors\\
$\ca{W}$ & the category of labelled boxes and wiring diagrams (with types in $\Set$) \\
$\ca{W}_\mathbf{Lin}$ & the category of labelled boxes and wiring diagrams (with types in $\mathbf{Lin}$) \\
$\mathcal{M}$ & the algebra of Moore machines; a lax monoidal functor $\ca{W}{\to}\mathbf{Cat}$\\
$\mathcal{L}$ & the algebra of linear time-invariant systems; a lax monoidal functor $\ca{W}_\mathbf{Lin}{\to}\mathbf{Cat}$\\
$\mathcal{C}$ & the algebra of (static) contracts\\
$\mathcal{B}$ & a generic behavior algebra; could be $\mathcal{M}$ or $\mathcal{L}$ (among others)\\
 \end{longtable}
 }
 
 \section{Contract Pullback}\label{sec:appendix}

Regarding the static contract algebra (\cref{sec:staticcontracts}), the functor $\mathcal{C}f\colon\mathcal{C}(X)\to\mathcal{C}(Y)$ for a wiring diagram $f\colon X\to Y$ \cref{eq:wirdiag}
assigns a contract $R_X\subseteq X_\inp\times X_\out$ on the inside box to a contract $R_X\subseteq X_\inp\times X_\out$ on the outside box, following a two-step procedure:
\begin{equation}\label{eq:pbdiagram}
\begin{tikzcd}[row sep=.5in, column sep=.5in]
& P\ar[dr,phantom,very near start,"{\scalebox{1.5}{$\lrcorner$}}"]\ar[dl,two heads]\ar[r]\ar[d,hook] & R_X\ar[d,hook] \\
R_Y\ar[dr,hook] & Y_\inp\times X_\out\ar[d,"{1\times f_\out}"]\ar[r,"{(f_\inp,\pi_2)}"'] & X_\inp\times X_\out \\
& Y_\inp\times Y_\out &
\end{tikzcd}
\end{equation}
First, we compute the pullback -- a limit of a diagram of two morphisms with common codomain~\cite[5.1.16]{leinster} -- of the relation $R_X$ along the function $(f_\inp,\pi_2)$ which is defined by $Y_\inp\times X_\out\ni (y,x')\mapsto (f_\inp(y,x'),x')\in X_\inp\times X_\out$. The explicit description of that pullback in $\Set$ is
\begin{displaymath}
P=\{(y,x')\;|\;(f_\inp(y,x'),x')\in R_X\}
\end{displaymath}
namely those pairs of $Y$-inputs and $X$-outputs which the bottom function actually maps to elements of the contract $R_X$. 
Second, we take the image factorization of the inclusion $P\subseteq Y_\inp\times X_\out$  post-composed with the function $1\times f_\out$ that maps some $(y,x')$ to the pair $(y,f_\out(x'))$. The image of a function is the subset of its codomain where all elements of the domain get mapped to, namely for an arbitrary $g\colon A\to B$, $\mathrm{Im}(g)=\{b\in B\;|\;\exists a\in A\textrm{ such that }g(a)=b\}$.
In the end, using the above constructions of the two-step process exhibited in \cref{eq:pbdiagram}, the explicit description of the resulting contract is precisely equation \cref{eq:compositecontract}.

\end{document}